\documentclass[prd,showkeys,floatfix,twocolumn,amsmath,amssymb,floatfix]{revtex4}
\usepackage{graphicx}
\usepackage{epstopdf}
\usepackage{multirow}
\usepackage{subfigure}
\usepackage[colorlinks,citecolor=blue,anchorcolor=red,menucolor=red,linkcolor=red,filecolor=red,runcolor=red,urlcolor=blue,frenchlinks=red]{hyperref}
\newcommand{\feynp}[1]{#1\kern-0.45em/}

\allowdisplaybreaks[3]

\begin{document}
\title{$P$-wave bottom baryons of the $SU(3)$ flavor $\mathbf{6}_F$}
%

\author{Hui-Min Yang$^1$}
\author{Hua-Xing Chen$^{1,2}$}
\email{hxchen@buaa.edu.cn}

\affiliation{
$^1$School of Physics, Beihang University, Beijing 100191, China \\
$^2$School of Physics, Southeast University, Nanjing 210094, China
}

\begin{abstract}
We investigate $P$-wave bottom baryons of the $SU(3)$ flavor $\mathbf{6}_F$, and systematically study their $D$-wave decays into ground-state bottom baryons and pseudoscalar mesons. Together with Refs.~\cite{Chen:2015kpa,Mao:2015gya,Chen:2017sci,Yang:2019cvw}, a rather complete study is performed on both mass spectra and decay properties of $P$-wave bottom baryons, using the method of QCD sum rules and light-cone sum rules within the framework of heavy quark effective theory. Among all the possibilities, we find four $\Sigma_b$, four $\Xi^\prime_b$, and six $\Omega_b$ baryons, with limited widths and so capable of being observed. Their masses, mass splittings within the same multiplets, and decay properties are extracted (summarized in Table~\ref{tab:result}) for future experimental searches.
\end{abstract}
\maketitle
\pagenumbering{arabic}
%
%
%
\section{Introduction}\label{sec:intro}
%

The strong interaction holds quarks and gluons together inside a single hadron. It is similar to the electromagnetic interaction in some aspects, which holds electrons and protons together inside a single atom. The latter leads to the well-known fine structure of line spectra, and it is interesting to investigate whether the former also leads to some fine structure of hadron spectra~\cite{Chen:2016spr,Copley:1979wj,Karliner:2008sv,pdg}. An ideal platform to study this is the singly bottom baryon system~\cite{Korner:1994nh,Manohar:2000dt,Bianco:2003vb,Klempt:2009pi}: light quarks together with gluons circle around the nearly static bottom quark, and the whole system behaves as the QCD analogue of the hydrogen.

In recent years important experimental progresses have been made in the field of singly bottom baryons. Until three years ago there were only two excited bottom baryons, $\Lambda_b(5912)^0$ and $\Lambda_b(5920)^0$, which were observed by LHCb and CDF in 2012~\cite{Aaij:2012da,Aaltonen:2013tta}. However, in the past three years the LHCb and CMS Collaborations discovered as many as nine excited bottom baryons:
\begin{itemize}

\item In 2018 the LHCb Collaboration observed the $\Sigma_{b}(6097)^{\pm}$ in the $\Lambda^0_b \pi^\pm$ invariant mass spectrum, and the $\Xi_{b}(6227)^{-}$ in the $\Lambda^0_b K^-$ and $\Xi^0_b \pi^-$ invariant mass spectra~\cite{Aaij:2018yqz,Aaij:2018tnn}:
    \begin{eqnarray}
    \nonumber              \Xi_{b}(6227)^{-}:M&=&6226.9 \pm 2.0 \pm 0.3 \pm 0.2 \mbox{ MeV} \, ,
    \\                         \Gamma&=&18.1 \pm 5.4 \pm 1.8 \mbox{ MeV} \, ,
    \\ \nonumber        \Sigma_{b}(6097)^{+}:M&=&6095.8 \pm 1.7 \pm 0.4 \mbox{ MeV} \, ,
    \\                         \Gamma&=&31 \pm 5.5 \pm 0.7 \mbox{ MeV} \, ,
    \\ \nonumber        \Sigma_{b}(6097)^{-}:M&=&6098.0 \pm 1.7 \pm 0.5 \mbox{ MeV} \, ,
    \\                                 \Gamma&=&28.9 \pm 4.2 \pm 0.9 \mbox{ MeV} \, .
    \end{eqnarray}

\item In 2020 the LHCb Collaboration observed the $\Omega_b(6316)^-$, $\Omega_b(6330)^-$, $\Omega_b(6340)^-$, and $\Omega_b(6350)^-$ in the $\Xi_b^0 K^-$ invariant mass spectrum~\cite{Aaij:2020cex}:
\begin{eqnarray}
\nonumber \Omega_b(6316)^-   &:& M = 6315.64 \pm 0.31 \pm 0.07 \pm 0.50~{\rm MeV} \, ,
\\                 && \Gamma < 2.8~{\rm MeV} \, ,
\\ \nonumber \Omega_b(6330)^-&:& M = 6330.30 \pm 0.28 \pm 0.07 \pm 0.50~{\rm MeV} \, ,
\\                 && \Gamma < 3.1 ~{\rm MeV} \, ,
\\ \nonumber \Omega_b(6340)^-&:& M = 6339.71 \pm 0.26 \pm 0.05 \pm 0.50~{\rm MeV} \, ,
\\                 && \Gamma < 1.5~{\rm MeV} \, ,
\\ \nonumber \Omega_b(6350)^-&:& M = 6349.88 \pm 0.35 \pm 0.05 \pm 0.50~{\rm MeV} \, ,
\\                 && \Gamma = 1.4^{+1.0}_{-0.8} \pm 0.1~{\rm MeV} \, .
\end{eqnarray}

\item In 2019 the LHCb Collaboration observed the $\Lambda_b(6146)^0$ and $\Lambda_b(6152)^0$ in the $\Lambda_b^0 \pi^+ \pi^-$ invariant mass distribution~\cite{Aaij:2019amv}. Later in 2020 the CMS Collaboration confirmed them, and further observed a broad excess of events in the $\Lambda_b^0 \pi^+ \pi^-$ mass distribution in the region of $6040$-$6100$~MeV~\cite{Sirunyan:2020gtz}, whose mass and width were later measured by LHCb to be~\cite{Aaij:2020rkw}:
    \begin{eqnarray}
    \nonumber \Lambda_b(6072)^0 &:& M = 6072.3 \pm 2.9 \pm 0.6 \pm 0.2~{\rm MeV} \, ,
    \\       && \Gamma = 72 \pm 11 \pm 2~{\rm MeV} \, .
    \end{eqnarray}

\end{itemize}

Various theoretical methods and models have been applied to study singly bottom baryons in the past thirty years, such as various quark models~\cite{Garcilazo:2007eh,Ebert:2007nw,Roberts:2007ni,Ortega:2012cx,Yoshida:2015tia,Nagahiro:2016nsx,Wang:2018fjm,Gutierrez-Guerrero:2019uwa,Kawakami:2019hpp,Xiao:2020oif}, various molecular models~\cite{GarciaRecio:2012db,Liang:2014eba,An:2017lwg,Montana:2017kjw,Debastiani:2017ewu,Chen:2017xat,Nieves:2017jjx,Huang:2018bed,Nieves:2019jhp,Liang:2020dxr,Huang:2018wgr,Yu:2018yxl}, the quark pair creation model~\cite{Chen:2018orb,Chen:2018vuc,Yang:2018lzg,Liang:2020hbo}, the chiral perturbation theory~\cite{Cheng:2006dk,Lu:2014ina,Cheng:2015naa},
QCD sum rules~\cite{Aliev:2018vye,Aliev:2018lcs,Wang:2020pri},
and
Lattice QCD~\cite{Padmanath:2013bla,Burch:2015pka,Padmanath:2017lng,Can:2019wts}, etc. More theoretical studies can be found in Refs.~\cite{Chen:2014nyo,Karliner:2015ema,Chua:2018lfa,Karliner:2018bms,Jia:2019bkr}, and we refer to recent reviews for detailed discussions~\cite{Chen:2016spr,Korner:1994nh,Manohar:2000dt,Bianco:2003vb,Klempt:2009pi,Crede:2013sze,Cheng:2015iom}.

Especially, the $\Lambda_b(5912)^0$ and $\Lambda_b(5920)^0$ were studied by Capstick and Isgur in 1986 as $P$-wave bottom baryons~\cite{Capstick:1986bm}, and their predicted masses are in very good agreement with the LHCb and CDF measurements~\cite{Aaij:2012da,Aaltonen:2013tta}. The $\Sigma_{b}(6097)^{\pm}$ and $\Xi_{b}(6227)^{-}$ are also good candidates of $P$-wave bottom baryons~\cite{Chen:2018orb,Chen:2018vuc,Yang:2018lzg,Wang:2018fjm,Aliev:2018vye,Aliev:2018lcs,Jia:2019bkr,Chua:2018lfa,Karliner:2018bms}, while there exists the molecular interpretation for the $\Xi_{b}(6227)^-$~\cite{Huang:2018bed,Yu:2018yxl}. Besides, the four excited $\Omega_b$ baryons, $\Omega_b(6316)^-$, $\Omega_b(6330)^-$, $\Omega_b(6340)^-$, and $\Omega_b(6350)^-$, are all good candidates of $P$-wave bottom baryons~\cite{Liang:2020hbo,Wang:2020pri,Xiao:2020oif}. Other than this, the $\Lambda_b(6146)^0$ and $\Lambda_b(6152)^0$ can be well interpreted as $D$-wave bottom baryons~\cite{Wang:2019uaj,Liang:2019aag,Chen:2019ywy,Azizi:2020tgh,Chen:2016phw,Mao:2017wbz,Mao:2020jln}.

In this paper we shall investigate $P$-wave bottom baryons of the $SU(3)$ flavor $\mathbf{6}_F$, and systematically study their $D$-wave decays into ground-state bottom baryons and pseudoscalar mesons. Previously in Refs.~\cite{Chen:2015kpa,Mao:2015gya,Chen:2017sci,Yang:2019cvw}, we have systematically studied their mass spectra and $S$-wave decay properties using the method of QCD sum rules~\cite{Shifman:1978bx,Reinders:1984sr} and light-cone sum rules~\cite{Balitsky:1989ry,Braun:1988qv,Chernyak:1990ag,Ball:1998je,Ball:2006wn} within the heavy quark effective theory (HQET)~\cite{Grinstein:1990mj,Eichten:1989zv,Falk:1990yz}. These results will be reanalysed in the present study, so that a rather complete study can be performed on both mass spectra and decay properties of $P$-wave bottom baryons. Similar methods applied to investigate singly heavy mesons and baryons can be found in Refs.~\cite{Bagan:1991sg,Neubert:1991sp,Broadhurst:1991fc,Huang:1994zj,Dai:1996yw,Colangelo:1998ga,Groote:1996em,Zhu:2000py,Lee:2000tb,Huang:2000tn,Wang:2003zp,Duraes:2007te,Liu:2007fg,Zhou:2014ytp,Zhou:2015ywa}.

This paper is organized as follows. In Sec.~\ref{sec:sumrule}, we briefly introduce our notations for $P$-wave bottom baryons of the $SU(3)$ flavor $\mathbf{6}_F$, and categorize them into four bottom baryon multiplets $[\mathbf{6}_F, 1, 0, \rho]$, $[\mathbf{6}_F, 0, 1, \lambda]$, $[\mathbf{6}_F, 1, 1, \lambda]$, and $[\mathbf{6}_F, 2, 1, \lambda]$. Then in Sec.~\ref{sec:decay} we study their $D$-wave decays into ground-state bottom baryons and pseudoscalar mesons ($\pi$ or $K$), separately for these four multiplets. In Sec.~\ref{sec:summary} we discuss the results and conclude this paper.

%
\section{$P$-wave bottom baryons}\label{sec:sumrule}
%

\begin{figure*}[hbtp]
\begin{center}
\scalebox{0.63}{\includegraphics{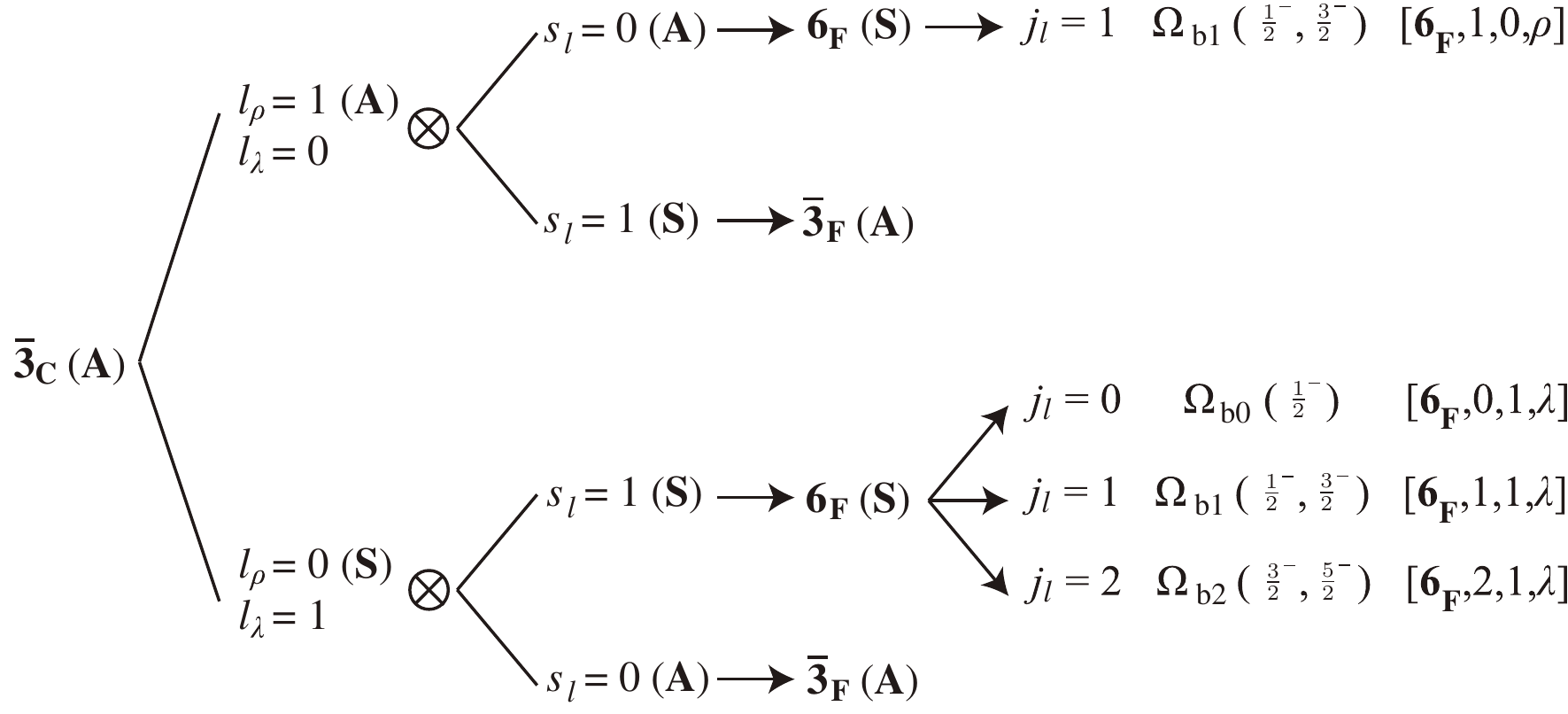}}
\end{center}
\caption{Categorization of $P$-wave bottom baryons belonging to the $SU(3)$ flavor $\mathbf{6}_F$ representation.}
\label{fig:pwave}
\end{figure*}

At the beginning we briefly introduce our notations. A singly bottom baryon is composed by one $bottom$ quark and two light $up/down/strange$ quarks, with the following internal structures:
\begin{itemize}

\item According to the Pauli principle, the total symmetry of the two light quarks is antisymmetric.

\item The color structure of the two light quarks is antisymmetric ($\mathbf{\bar 3}_C$).

\item The flavor structure of the two light quarks is either symmetric ($SU(3)$ flavor $\mathbf{6}_F$) or antisymmetric ($SU(3)$ flavor $\mathbf{\bar 3}_F$).

\item The spin structure of the two light quarks is either symmetric ($s_l \equiv s_{qq} = 1$) or antisymmetric ($s_l = 0$).

\item The orbital structure of the two light quarks is either symmetric ($\lambda$-type with $l_\rho = 0$ and $l_\lambda = 1$, meaning that the orbital excitation is between the bottom quark and the two-light-quark system) or antisymmetric ($\rho$-type with $l_\rho = 1$ and $l_\lambda = 0$, meaning that the orbital excitation is between the two light quarks).

\end{itemize}
Accordingly, we categorize $P$-wave bottom baryons into eight baryon multiplets, four of which belong to the $SU(3)$ flavor $\mathbf{6}_F$ representation, as shown in Fig.~\ref{fig:pwave}. We use $[F({\rm flavor}), j_l, s_l, \rho/\lambda]$ to denote them, where $j_l$ is the total angular momentum of the light components, satisfying $j_l = l_\lambda \otimes l_\rho \otimes s_l$. Each multiplet contains one or two bottom baryons with the total angular momenta $j = j_l \otimes s_b = | j_l \pm 1/2 |$, which have similar masses according to the heavy quark effective theory.

\begin{table*}[hbtp]
\begin{center}
\renewcommand{\arraystretch}{1.4}
\caption{Parameters of the $P$-wave bottom baryons belonging to the $SU(3)$ flavor $\mathbf{6}_F$ representation. See Refs.~\cite{Chen:2015kpa,Mao:2015gya} for detailed discussions. In the last column we list the decay constant $f$, where isospin factors are explicitly taken into account, satisfying $f_{\Sigma^+_b} = f_{\Sigma^-_b} = \sqrt2 f_{\Sigma^0_b}$ and $f_{\Xi^{\prime0}_b} = f_{\Xi^{\prime-}_b}$.}
\begin{tabular}{c | c | c | c | c | c c | c | c}
\hline\hline
\multirow{2}{*}{Multiplets} & \multirow{2}{*}{~B~} & $\omega_c$ & ~~~Working region~~~ & ~~~~~~~$\overline{\Lambda}$~~~~~~~ & ~~~Baryons~~~ & ~~~Mass~~~ & ~Difference~ & $f$
\\                                              &  & (GeV)      & (GeV)                & (GeV)                              & ($j^P$)       & (GeV)      & (MeV)        & (GeV$^{4}$)
\\ \hline\hline
\multirow{6}{*}{$[\mathbf{6}_F, 1, 0, \rho]$}
& \multirow{2}{*}{$\Sigma_b$} & \multirow{2}{*}{1.83} & \multirow{2}{*}{$0.27< T < 0.34$} & \multirow{2}{*}{$1.31 \pm 0.11$} & $\Sigma_b(1/2^-)$ & $6.05 \pm 0.12$ & \multirow{2}{*}{$3 \pm 1$} & $0.079 \pm 0.019~(\Sigma^-_b(1/2^-))$
\\ \cline{6-7}\cline{9-9}
& & & & & $\Sigma_b(3/2^-)$ & $6.05 \pm 0.12$ & &$0.037 \pm 0.009~(\Sigma^-_b(3/2^-))$
\\ \cline{2-9}
& \multirow{2}{*}{$\Xi^\prime_b$} & \multirow{2}{*}{1.98} & \multirow{2}{*}{$0.26< T < 0.36$} & \multirow{2}{*}{$1.45 \pm 0.11$} & $\Xi^\prime_b(1/2^-)$ & $6.18 \pm 0.12$ & \multirow{2}{*}{$3 \pm 1$} & $0.072 \pm 0.016~(\Xi^{\prime-}_b(1/2^-))$
\\ \cline{6-7}\cline{9-9}
& & & & & $\Xi^\prime_b(3/2^-)$ & $6.19 \pm 0.11$ & &$0.034 \pm 0.008~(\Xi^{\prime-}_b(3/2^-))$
\\ \cline{2-9}
& \multirow{2}{*}{$\Omega_b$} & \multirow{2}{*}{2.13} & \multirow{2}{*}{$0.26< T < 0.37$} & \multirow{2}{*}{$1.58 \pm 0.09$} & $\Omega_b(1/2^-)$ & $6.32 \pm 0.11$ & \multirow{2}{*}{$2 \pm 1$} & $0.133 \pm 0.028~(\Omega^-_b(1/2^-))$
\\ \cline{6-7}\cline{9-9}
& & & & & $\Omega_b(3/2^-)$ & $6.32 \pm 0.11$ & &$0.063 \pm 0.013~(\Omega^-_b(3/2^-))$
\\ \hline
\multirow{3}{*}{$[\mathbf{6}_F, 0, 1, \lambda]$} & $\Sigma_b$ & $1.70$ & $0.26< T < 0.32$ & $1.25 \pm 0.10$ & $\Sigma_b(1/2^-)$ & $6.05 \pm 0.11$ & -- & $0.077 \pm 0.018~(\Sigma^-_b(1/2^-))$
\\ \cline{2-9}
                                                 & $\Xi^\prime_b$ & $1.85$ & $0.27< T < 0.33$ & $1.40 \pm 0.09$ & $\Xi^\prime_b(1/2^-)$ & $6.20 \pm 0.11$ & -- & $0.069 \pm 0.015~(\Xi^{\prime-}_b(1/2^-))$
\\ \cline{2-9}
                                                 & $\Omega_b$ & 2.00 & $0.27< T < 0.34$ & $1.54 \pm 0.09$ & $\Omega_b(1/2^-)$ & $6.34 \pm 0.11$ & -- & $0.127 \pm 0.028~(\Omega^-_b(1/2^-))$
\\ \hline
\multirow{6}{*}{$[\mathbf{6}_F, 1, 1, \lambda]$}
& \multirow{2}{*}{$\Sigma_b$} & \multirow{2}{*}{1.94} & \multirow{2}{*}{$0.29< T < 0.36$} & \multirow{2}{*}{$1.25 \pm 0.11$} & $\Sigma_b(1/2^-)$ & $6.06 \pm 0.13$ & \multirow{2}{*}{$6 \pm 3$} & $0.075 \pm 0.016~(\Sigma^-_b(1/2^-))$
\\ \cline{6-7}\cline{9-9}
& & & & & $\Sigma_b(3/2^-)$ & $6.07 \pm 0.13$ & &$0.035 \pm 0.008~(\Sigma^-_b(3/2^-))$
\\ \cline{2-9}
& \multirow{2}{*}{$\Xi^\prime_b$} & \multirow{2}{*}{1.97} & \multirow{2}{*}{$0.35< T < 0.38$} & \multirow{2}{*}{$1.38 \pm 0.09$} & $\Xi^\prime_b(1/2^-)$ & $6.21 \pm 0.11$ & \multirow{2}{*}{$7 \pm 2$} & $0.069 \pm 0.012~(\Xi^{\prime-}_b(1/2^-))$
\\ \cline{6-7}\cline{9-9}
& & & & & $\Xi^\prime_b(3/2^-)$ & $6.22 \pm 0.11$ & &$0.032 \pm 0.006~(\Xi^{\prime-}_b(3/2^-))$
\\ \cline{2-9}
& \multirow{2}{*}{$\Omega_b$} & \multirow{2}{*}{2.00} & \multirow{2}{*}{$0.38< T < 0.39$} & \multirow{2}{*}{$1.48 \pm 0.07$} & $\Omega_b(1/2^-)$ & $6.34 \pm 0.10$ & \multirow{2}{*}{$6 \pm 2$} & $0.122 \pm 0.019~(\Omega^-_b(1/2^-))$
\\ \cline{6-7}\cline{9-9}
& & & & & $\Omega_b(3/2^-)$ & $6.34 \pm 0.09$ & &$0.058 \pm 0.009~(\Omega^-_b(3/2^-))$
\\ \hline
\multirow{6}{*}{$[\mathbf{6}_F, 2, 1, \lambda]$}
& \multirow{2}{*}{$\Sigma_b$} & \multirow{2}{*}{1.84} & \multirow{2}{*}{$0.27< T < 0.34$} & \multirow{2}{*}{$1.30 \pm 0.13$} & $\Sigma_b(3/2^-)$ & $6.11 \pm 0.16$ & \multirow{2}{*}{$12 \pm 5$} & $0.102 \pm 0.028~(\Sigma^-_b(3/2^-))$
\\ \cline{6-7}\cline{9-9}
& & & & & $\Sigma_b(5/2^-)$ & $6.12 \pm 0.15$ & &$0.061 \pm 0.016~(\Sigma^-_b(5/2^-))$
\\ \cline{2-9}
 & \multirow{2}{*}{$\Xi^\prime_b$} & \multirow{2}{*}{1.96} & \multirow{2}{*}{$0.26< T < 0.35$} & \multirow{2}{*}{$1.41 \pm 0.12$} & $\Xi^\prime_b(3/2^-)$ & $6.23 \pm 0.15$ & \multirow{2}{*}{$11 \pm 5$} & $0.091 \pm 0.023~(\Xi^{\prime-}_b(3/2^-))$
\\ \cline{6-7}\cline{9-9}
& & & & & $\Xi^\prime_b(5/2^-)$ & $6.24 \pm 0.14$ & &$0.054 \pm 0.013~(\Xi^{\prime-}_b(5/2^-))$
\\ \cline{2-9}
& \multirow{2}{*}{$\Omega_b$} & \multirow{2}{*}{2.08} & \multirow{2}{*}{$0.26< T < 0.37$} & \multirow{2}{*}{$1.53 \pm 0.10$} & $\Omega_b(3/2^-)$ & $6.35 \pm 0.13$ & \multirow{2}{*}{$10 \pm 4$} & $0.162 \pm 0.035~(\Omega^-_b(3/2^-))$
\\ \cline{6-7}\cline{9-9}
& & & & & $\Omega_b(5/2^-)$ & $6.36 \pm 0.12$ & &$0.097 \pm 0.021~(\Omega^-_b(5/2^-))$
\\ \hline \hline
\\
\end{tabular}
\label{tab:pwaveparameter}
\end{center}
\end{table*}

In Refs.~\cite{Chen:2015kpa,Mao:2015gya} we have systematically studied the mass spectrum of $P$-wave bottom baryons, and the results are reanalysed in the present study, as summarized in Table~\ref{tab:pwaveparameter}. Some of them are used as input parameters when studying decay properties of $P$-wave bottom baryons. Especially, we use the following mass values when calculating their decay widths:
\begin{itemize}

\item In Ref.~\cite{Chen:2020mpy} we found that the $\Omega_b(6316)^-$ can be explained as a $P$-wave $\Omega_b$ baryon of either $J^P = 1/2^-$ or $3/2^-$, belonging to the $[\mathbf{6}_F, 1, 0, \rho]$ doublet. Hence, we use the following mass values for this doublet, taken from the LHCb experiment~\cite{Aaij:2020cex} as well as their mass sum rules:
\begin{eqnarray}
\nonumber M_{[\Sigma_b(1/2^-), 1, 0, \rho]} &=& 6.05~{\rm GeV} \, ,
\\ \nonumber M_{[\Sigma_b(3/2^-), 1, 0, \rho]} &=& 6.05~{\rm GeV} \, ,
\\ M_{[\Xi_b^{\prime}(1/2^-), 1, 0, \rho]} &=& 6.18~{\rm GeV} \, ,
\\ \nonumber M_{[\Xi_b^{\prime}(3/2^-), 1, 0, \rho]} &=& 6.19~{\rm GeV} \, ,
\\ \nonumber M_{[\Omega_b(1/2^-), 1, 0, \rho]} &=& 6315.64~{\rm MeV} \, ,
\\ \nonumber M_{[\Omega_b(3/2^-), 1, 0, \rho]} &=& 6315.64~{\rm MeV} \, .
\end{eqnarray}

\item For the $[\mathbf{6}_F, 0, 1, \lambda]$ singlet, we use the following mass values taken from their mass sum rules:
\begin{eqnarray}
\nonumber M_{[\Sigma_b(1/2^-), 0, 1, \lambda]} &=& 6.05~{\rm GeV}\, ,
\\ M_{[\Xi_b^{\prime}(1/2^-), 0, 1, \lambda]} &=& 6.20~{\rm GeV}\, ,
\\ \nonumber M_{[\Omega_b(1/2^-), 0, 1, \lambda]} &=& 6.34~{\rm GeV} \, .
\end{eqnarray}

\item In Ref.~\cite{Chen:2020mpy} we found that the $\Omega_b(6330)^-$ and $\Omega_b(6340)^-$ can be explained as $P$-wave $\Omega_b$ baryons of $J^P = 1/2^-$ and $3/2^-$ respectively, belonging to the $[\mathbf{6}_F, 1, 1, \lambda]$ doublet. Hence, we use the following mass values for this doublet, taken from the LHCb experiment~\cite{Aaij:2020cex} as well as their mass sum rules:
\begin{eqnarray}
\nonumber M_{[\Sigma_b(1/2^-), 1, 1, \lambda]} &=& 6.06~{\rm GeV} \, ,
\\ \nonumber M_{[\Sigma_b(3/2^-), 1, 1, \lambda]} &=& 6.07~{\rm GeV} \, ,
\\ M_{[\Xi_b^{\prime}(1/2^-), 1, 1, \lambda]} &=& 6.21~{\rm GeV} \, ,
\\ \nonumber M_{[\Xi_b^{\prime}(3/2^-), 1, 1, \lambda]} &=& 6.22~{\rm GeV} \, ,
\\ \nonumber M_{[\Omega_b(1/2^-), 1, 1, \lambda]} &=& 6330.30~{\rm MeV} \, ,
\\ \nonumber M_{[\Omega_b(3/2^-), 1, 1, \lambda]} &=& 6339.71~{\rm MeV} \, .
\end{eqnarray}

\item In Refs.~\cite{Cui:2019dzj,Chen:2020mpy} we found that the $\Sigma_{b}(6097)^{\pm}$, $\Xi_{b}(6227)^{-}$, and $\Omega_b(6350)^-$ can be explained as $P$-wave bottom baryons of $J^P = 3/2^-$, belonging to the $[\mathbf{6}_F, 2, 1, \lambda]$ doublet. Hence, we use the following mass values for this doublet, taken from the LHCb experiments~\cite{Aaij:2018yqz,Aaij:2018tnn,Aaij:2020cex} as well as their mass sum rules:
\begin{eqnarray}
\nonumber M_{[\Sigma_b(3/2^-), 2, 1, \lambda]} &=& 6096.9~{\rm MeV}\, ,
\\ \nonumber M_{[\Sigma_b(5/2^-), 2, 1, \lambda]} -  M_{[\Sigma_b(3/2^-), 2, 1, \lambda]} &=& 12~{\rm MeV} \, ,
\\ \nonumber M_{[\Xi_b^{\prime}(3/2^-), 2, 1, \lambda]} &=& 6226.9~{\rm MeV}\, ,
\\ \nonumber M_{[\Xi_b^{\prime}(5/2^-), 2, 1, \lambda]} -  M_{[\Xi_b^{\prime}(3/2^-), 2, 1, \lambda]} &=& 11~{\rm MeV} \, ,
\\ \nonumber M_{[\Omega_b(3/2^-), 2, 1, \lambda]} &=& 6349.88~{\rm MeV} \, ,
\\ M_{[\Omega_b(5/2^-), 2, 1, \lambda]} -  M_{[\Omega_b(3/2^-), 2, 1, \lambda]} &=& 10~{\rm MeV} \, .
\end{eqnarray}

\end{itemize}
Note that the above interpretations are just possible assignments, and there exist many other possibilities for the $\Sigma_{b}(6097)^{\pm}$, $\Xi_{b}(6227)^{-}$, $\Omega_b(6316)^-$, $\Omega_b(6330)^-$, $\Omega_b(6340)^-$, and $\Omega_b(6350)^-$.

We use the following parameters for ground-state bottom baryons, pseudoscalar and vector mesons~\cite{pdg}:
\begin{eqnarray}
   \nonumber        \Lambda_{b}(1/2^+)  ~:~ m&=&5619.60 \mbox{ MeV} \, ,
\\ \nonumber        \Xi_{b}(1/2^+)  ~:~ m&=&5793.20 \mbox{ MeV} \, ,
\\ \nonumber        \Sigma_{b}(1/2^+)    ~:~ m&=&5813.4 \mbox{ MeV} \, ,
\\ \nonumber        \Sigma_{b}^{*}(3/2^+)    ~:~ m&=&5833.6 \mbox{ MeV} \, , \,
\\ \nonumber        \Xi_{b}^{\prime}(1/2^+)  ~:~ m&=&5935.02 \mbox{ MeV} \, ,
\\ \nonumber        \Xi_{b}^{*}(3/2^+)  ~:~ m&=&5952.6 \mbox{ MeV} \, , \,
\\         \Omega_b(1/2^+)  ~:~ m&=&6046.1 \mbox{ MeV}\, ,
\\ \nonumber        \Omega_b^*(3/2^+)  ~:~ m&=&6063 \mbox{ MeV} \, ,
\\ \nonumber  \pi(0^-) ~:~ m&=& 138.04 {\rm~MeV} \, ,
\\ \nonumber K(0^-) ~:~ m&=& 495.65 {\rm~MeV} \, ,
\\ \nonumber \rho(1^-) ~:~ m&=& 775.21 {\rm~MeV} \, ,\,
\\ \nonumber                                \Gamma&=& 148.2 {\rm~MeV} \, ,\,  {g}_{\rho \pi \pi} = 5.94 \, ,
\\  \nonumber         K^*(1^-) ~:~ m&=& 893.57 {\rm~MeV} \, ,\,
\\   \nonumber                 \Gamma&=& 49.1 {\rm~MeV} \, ,\,  {g}_{K^* K \pi} = 3.20 \, ,
\end{eqnarray}
with the following Lagrangians:
\begin{eqnarray}
\nonumber \mathcal{L}_{\rho \pi \pi} &=& {g}_{\rho \pi \pi} \times \left( \rho_\mu^0 \pi^+ \partial^\mu \pi^- - \rho_\mu^0 \pi^- \partial^\mu \pi^+ \right) + \cdots \, ,
\\ \nonumber \mathcal{L}_{K^* K \pi} &=& {g}_{K^* K \pi} K^{*+}_\mu \times \left( K^- \partial^\mu \pi^0 - \partial^\mu K^- \pi^0 \right) + \cdots \, .
\\
\end{eqnarray}

\section{$D$-wave decay properties}\label{sec:decay}

In the present study we shall investigate $P$-wave bottom baryons of the $SU(3)$ flavor $\mathbf{6}_F$, and study their $D$-wave decays into ground-state bottom baryons together with pseudoscalar mesons ($\pi$ or $K$). Their $S$-wave decay properties have been systematically studied in Refs.~\cite{Chen:2017sci,Yang:2019cvw}, and we shall use the same method of light-cone sum rules within the heavy quark effective theory to investigate the following decay channels (the coefficients at right hand sides are isospin factors):
\begin{widetext}
\begin{eqnarray}
&(a1)& {\bf \Gamma\Big[} \Sigma_b[1/2^-] \rightarrow \Sigma_b^{*} + \pi {\Big ]}
= 2 \times {\bf \Gamma\Big[} \Sigma_b^{-}[1/2^-] \rightarrow \Sigma_b^{*0} +\pi^- {\Big ]} \, ,
\label{eq:couple1}
\\ &(b1)& {\bf \Gamma\Big[}\Sigma_b[3/2^-] \rightarrow \Lambda_b + \pi{\Big ]}
= {\bf \Gamma\Big[}\Sigma_b^{-}[3/2^-] \rightarrow \Lambda_b^{0} +\pi^-{\Big ]} \, ,
\\ &(b2)&{\bf \Gamma\Big[}\Sigma_b[3/2^-] \rightarrow \Sigma_b + \pi{\Big ]}
= 2 \times {\bf \Gamma\Big[}\Sigma_b^{-}[3/2^-] \rightarrow \Sigma_b^{0} +\pi^-{\Big ]} \, ,
\\ &(b3)&{\bf \Gamma\Big[}\Sigma_b[3/2^-] \rightarrow \Sigma_b^{*} + \pi{\Big ]}
= 2 \times {\bf \Gamma\Big[}\Sigma_b^{-}[3/2^-] \rightarrow \Sigma_b^{*0} +\pi^-{\Big ]} \, ,
\\ &(c1)& {\bf \Gamma\Big[}\Sigma_b[5/2^-] \rightarrow \Lambda_b + \pi{\Big ]}
= {\bf \Gamma\Big[}\Sigma_b^{-}[5/2^-] \rightarrow \Lambda_b^{0} +\pi^-{\Big ]} \, ,
\\ &(c2)&{\bf \Gamma\Big[}\Sigma_b[5/2^-] \rightarrow \Sigma_b + \pi{\Big ]}
= 2 \times {\bf \Gamma\Big[}\Sigma_b^{-}[5/2^-] \rightarrow \Sigma_b^{0} +\pi^-{\Big ]} \, ,
\\ &(c3)&{\bf \Gamma\Big[}\Sigma_b[5/2^-] \rightarrow \Sigma_b^{*} + \pi{\Big ]}
= 2 \times {\bf \Gamma\Big[}\Sigma_b^{-}[5/2^-] \rightarrow \Sigma_b^{*0} +\pi^-{\Big ]} \, ,
\\ &(d1)& {\bf \Gamma\Big[}\Xi_b^{\prime}[1/2^-] \rightarrow \Xi_b^{*} + \pi{\Big ]}
= {3 \over 2} \times {\bf \Gamma\Big[}\Xi_b^{\prime -}[1/2^-] \rightarrow \Xi_b^{*0} + \pi^-{\Big ]} \, ,
\\ &(d2)& {\bf \Gamma\Big[} \Xi_b^{\prime}[1/2^-] \rightarrow \Sigma_b^{*} + K {\Big ]}
= 3 \times {\bf \Gamma\Big[} \Xi_b^{\prime -}[1/2^-] \rightarrow \Sigma_b^{*0} + K^- {\Big ]} \, ,
\\ &(e1)& {\bf \Gamma\Big[}\Xi_b^{\prime}[3/2^-] \rightarrow \Xi_b + \pi {\Big ]}
= {3 \over 2} \times  {\bf \Gamma\Big[}\Xi_b^{\prime -}[3/2^-] \rightarrow \Xi_b^{0} + \pi^- {\Big ]} \, ,
\\ &(e2)& {\bf \Gamma\Big[}\Xi_b^{\prime}[3/2^-] \rightarrow \Lambda_b + K {\Big ]}
= {\bf \Gamma\Big[}\Xi_b^{\prime -}[3/2^-] \rightarrow \Lambda_b^{0} + K^- {\Big ]} \, ,
\\&(e3)&{\bf \Gamma\Big[}\Xi_b^{\prime}[3/2^-] \rightarrow \Xi_b^{\prime} + \pi {\Big ]}
= {3 \over 2} \times  {\bf \Gamma\Big[}\Xi_b^{\prime -}[1/2^-] \rightarrow \Xi_b^{\prime 0} + \pi^- {\Big ]} \, ,
\\&(e4)&{\bf \Gamma\Big[}\Xi_b^{\prime}[3/2^-] \rightarrow \Sigma_b + K {\Big ]}
= 3 \times  {\bf \Gamma\Big[}\Xi_b^{\prime -}[3/2^-] \rightarrow \Sigma_b^{0} + K^- {\Big ]} \, ,
\\ &(e5)& {\bf \Gamma\Big[}\Xi_b^{\prime}[3/2^-] \rightarrow \Xi_b^{*} + \pi {\Big ]}
= {3 \over 2} \times {\bf \Gamma\Big[}\Xi_b^{\prime-}[3/2^-] \rightarrow \Xi_b^{*0} + \pi^- {\Big ]} \, ,
\\ &(e6)& {\bf \Gamma\Big[}\Xi_b^{\prime}[3/2^-] \rightarrow \Sigma_b^{*} + K {\Big ]}
= 3 \times {\bf \Gamma\Big[}\Xi_b^{\prime-}[3/2^-] \rightarrow \Sigma_b^{*0} + K^- {\Big ]} \, ,
\\ &(f1)& {\bf \Gamma\Big[}\Xi_b^{\prime}[5/2^-] \rightarrow \Xi_b + \pi {\Big ]}
= {3 \over 2} \times  {\bf \Gamma\Big[}\Xi_b^{\prime -}[5/2^-] \rightarrow \Xi_b^{0} + \pi^- {\Big ]} \, ,
\\ &(f2)& {\bf \Gamma\Big[}\Xi_b^{\prime}[5/2^-] \rightarrow \Lambda_b + K {\Big ]}
= {\bf \Gamma\Big[}\Xi_b^{\prime -}[5/2^-] \rightarrow \Lambda_b^{0} + K^- {\Big ]} \, ,
\\&(f3)&{\bf \Gamma\Big[}\Xi_b^{\prime}[5/2^-] \rightarrow \Xi_b^{\prime} + \pi {\Big ]}
= {3 \over 2} \times  {\bf \Gamma\Big[}\Xi_b^{\prime -}[1/2^-] \rightarrow \Xi_b^{\prime 0} + \pi^- {\Big ]} \, ,
\\&(f4)&{\bf \Gamma\Big[}\Xi_b^{\prime}[5/2^-] \rightarrow \Sigma_b + K {\Big ]}
= 3 \times  {\bf \Gamma\Big[}\Xi_b^{\prime -}[5/2^-] \rightarrow \Sigma_b^{0} + K^- {\Big ]} \, ,
\\ &(f5)& {\bf \Gamma\Big[}\Xi_b^{\prime}[5/2^-] \rightarrow \Xi_b^{*} + \pi {\Big ]}
= {3 \over 2} \times {\bf \Gamma\Big[}\Xi_b^{\prime-}[5/2^-] \rightarrow \Xi_b^{*0} + \pi^- {\Big ]} \, ,
\\ &(f6)& {\bf \Gamma\Big[}\Xi_b^{\prime}[5/2^-] \rightarrow \Sigma_b^{*} + K {\Big ]}
= 3 \times {\bf \Gamma\Big[}\Xi_b^{\prime-}[5/2^-] \rightarrow \Sigma_b^{*0} + K^- {\Big ]} \, ,
\\ &(g1)&{\bf \Gamma\Big[}\Omega_b[1/2^-] \rightarrow \Xi_b^{*} + K {\Big ]}
= 2 \times {\bf \Gamma\Big[}\Omega_b^{-}[1/2^-] \rightarrow \Xi_b^{*0} + K^- {\Big ]} \, ,
\\ &(h1)&{\bf \Gamma\Big[} \Omega_b[3/2^-] \rightarrow \Xi_b + K {\Big ]}
= 2 \times {\bf \Gamma\Big[} \Omega_b^{-}[3/2^-] \rightarrow \Xi_b^{0} +k^- {\Big ]} \, ,
\\ &(h2)& {\bf \Gamma\Big[}\Omega_b[3/2^-] \rightarrow \Xi_b^{\prime} + K {\Big ]}
= 2 \times {\bf \Gamma\Big[}\Omega_b^{-}[3/2^-] \rightarrow \Xi_b^{\prime0} + K^- {\Big ]} \, ,
\\ &(h3)& {\bf \Gamma\Big[}\Omega_b[3/2^-] \rightarrow \Xi_b^{*} + K {\Big ]}
= 2 \times {\bf \Gamma\Big[}\Omega_b^{-}[3/2^-] \rightarrow \Xi_b^{*0} + K^- {\Big ]} \, ,
\\ &(i1)&{\bf \Gamma\Big[} \Omega_b[5/2^-] \rightarrow \Xi_b + K {\Big ]}
= 2 \times {\bf \Gamma\Big[} \Omega_b^{-}[5/2^-] \rightarrow \Xi_b^{0} +k^- {\Big ]} \, ,
\\ &(i2)& {\bf \Gamma\Big[}\Omega_b[5/2^-] \rightarrow \Xi_b^{\prime} + K {\Big ]}
= 2 \times {\bf \Gamma\Big[}\Omega_b^{-}[5/2^-] \rightarrow \Xi_b^{\prime0} + K^- {\Big ]} \, ,
\\ &(i3)& {\bf \Gamma\Big[}\Omega_b[5/2^-] \rightarrow \Xi_b^{*} + K {\Big ]}
= 2 \times {\bf \Gamma\Big[}\Omega_b^{-}[5/2^-] \rightarrow \Xi_b^{*0} + K^- {\Big ]} \, .
\label{eq:couple28}
\end{eqnarray}
\end{widetext}
We shall calculate their decay widths through:
\begin{eqnarray}
&& \mathcal{L}_{X_b({1/2}^-) \rightarrow Y_b({3/2}^+) P}
\\ \nonumber && ~~~~~~~~~~~\, = g {\bar X_b}(1/2^-) \gamma_\mu \gamma_5 Y_{b\nu}(3/2^+) \partial^{\mu} \partial^{\nu}P \, ,
\\ && \mathcal{L}_{X_b({3/2}^-) \rightarrow Y_b({1/2}^+) P}
\\ \nonumber && ~~~~~~~~~~~\, = g {\bar X_{b\mu}}(3/2^-) \gamma_\nu \gamma_5 Y_{b}(1/2^+) \partial^{\mu} \partial^{\nu}P \, ,
\\ && \mathcal{L}_{X_b({3/2}^-) \rightarrow Y_b({3/2}^+) P}
\\ \nonumber && ~~~~~~~~~~~\, = g {\bar X_{b\mu}}(3/2^-) Y_{b\nu}(3/2^+) \partial^{\mu} \partial^{\nu}P \, ,
\\ && \mathcal{L}_{X_b({5/2}^-) \rightarrow Y_b({1/2}^+) P}
\\ \nonumber && ~~~~~~~~~~~\, = g {\bar X_{b\mu\nu}}(5/2^-) Y_{b}(1/2^+) \partial^{\mu} \partial^{\nu}P \, ,
\\ && \mathcal{L}_{X_b({5/2}^-) \rightarrow Y_b({3/2}^+) P}
\\ \nonumber && ~~~~~~~~~~~\, = g {\bar X_{b\mu\nu}}(5/2^-) \gamma_\rho \gamma_5 Y_{b}^{\mu}(3/2^+) \partial^{\nu} \partial^{\rho}P
\\ \nonumber && ~~~~~~~~~~~\, + g {\bar X_{b\mu\nu}}(5/2^-) \gamma_\rho \gamma_5 Y_{b}^{\nu}(3/2^+) \partial^{\mu} \partial^{\rho}P \, ,
\end{eqnarray}
where $X_b^{(\mu\nu)}$, $Y_b^{(\mu)}$, and $P$ denote the $P$-wave bottom baryon, ground-state bottom baryon, and pseudoscalar meson, respectively.

Especially, we need the following expression for the $J=5/2$ propagator~\cite{Vrancx:2011qv}:
\begin{eqnarray}
&& P_{\mu\nu;\lambda\rho}(p)
\label{propagator}
\\ \nonumber &=& \frac{\feynp{p} + m}{p^2 - m^2}\biggl[\frac{1}{2}(g_{\mu\lambda}g_{\nu\rho}+g_{\mu\rho}g_{\nu\lambda}) - \frac{1}{5}g_{\mu\nu}g_{\lambda\rho}
\\ \nonumber && - \frac{1}{10}(g_{\mu\lambda}\gamma_\nu\gamma_\rho + g_{\mu\rho}\gamma_\nu\gamma_\lambda + g_{\nu\lambda}\gamma_\mu\gamma_\rho + g_{\nu\rho}\gamma_\mu\gamma_\lambda)
\\ \nonumber && + \frac{1}{10m}\Bigl(g_{\mu\lambda}(p_\nu\gamma_\rho - p_\rho\gamma_\nu) + g_{\mu\rho}(p_\nu\gamma_\lambda - p_\lambda\gamma_\nu)
\\ \nonumber && ~~~~~~~~~~~ + g_{\nu\lambda}(p_\mu\gamma_\rho - p_\rho\gamma_\mu) + g_{\nu\rho}(p_\mu\gamma_\lambda - p_\lambda\gamma_\mu)\Bigr)
\\ \nonumber && + \frac{1}{5m^2}(g_{\mu\nu}p_\lambda p_\rho + g_{\lambda\rho}p_\mu p_\nu)
\\ \nonumber && - \frac{2}{5m^2}(g_{\mu\lambda}p_\nu p_\rho + g_{\mu\rho}p_\nu p_\lambda + g_{\nu\lambda}p_\mu p_\rho + g_{\nu\rho}p_\mu p_\lambda)
\\ \nonumber && + \frac{1}{10m^2}(\gamma_\mu\gamma_\lambda p_\nu p_\rho + \gamma_\mu\gamma_\rho p_\nu p_\lambda
\\ \nonumber && ~~~~~~~~~~~ + \gamma_\nu\gamma_\lambda p_\mu p_\rho + \gamma_\nu\gamma_\rho p_\mu p_\lambda)
\\ \nonumber && + \frac{1}{5m^3}(\gamma_\mu p_\nu p_\lambda p_\rho + \gamma_\nu p_\mu p_\lambda p_\rho
\\ \nonumber && ~~~~~~~ - \gamma_\lambda p_\mu p_\nu p_\rho - \gamma_\rho p_\mu p_\nu p_\lambda) + \frac{2}{5m^4}p_\mu p_\nu p_\lambda p_\rho\biggr] \, .
\end{eqnarray}

As an example, we shall study the $D$-wave decay of $\Omega_b^-({3/2}^-)$ belonging to $[\mathbf{6}_F, 2, 1, \lambda]$ into $\Xi_b^0(1/2^+)$ and $K^-(0^-)$ in the next subsection. The four bottom baryon multiplets $[\mathbf{6}_F, 1, 0, \rho]$, $[\mathbf{6}_F, 0, 1, \lambda]$, $[\mathbf{6}_F, 1, 1, \lambda]$, and $[\mathbf{6}_F, 2, 1, \lambda]$ will be separately investigated in the following subsections.

\subsection{$\Omega_b^-({3/2}^-)$ of $[\mathbf{6}_F, 2, 1, \lambda]$ to $D$-wave $\Xi_b^0 K^-$}

In this subsection we study the $D$-wave decay of the $\Omega_b^-({3/2}^-)$ belonging to $[\mathbf{6}_F, 2, 1, \lambda]$ into $\Xi_b^0(1/2^+)$ and $K^-(0^-)$. To do this we need to calculate the three-point correlation function
\begin{eqnarray}
&& \Pi^\alpha(\omega, \omega^\prime)
\\ \nonumber &=& \int d^4 x~e^{-i k \cdot x}~\langle 0 | J^\alpha_{3/2,-,\Omega_b^-,2,1,\lambda}(0) \bar J_{\Xi_b^{0}}(x) | K^-(q) \rangle
\\ \nonumber &=& {1+v\!\!\!\slash\over2} G^\alpha_{\Omega_b^-[{3\over2}^-] \rightarrow   \Xi_b^{0}K^-} (\omega, \omega^\prime) \, .
\end{eqnarray}
at both hadron and quark-gluon levels. In this expression $k^\prime = k + q$, $\omega = v \cdot k$, and $\omega^\prime = v \cdot k^\prime$. The two interpolating fields $J^\alpha_{3/2,-,\Omega_b^-,2,1,\lambda}$ and $J_{\Xi_b^{0}}$ have been constructed in Refs.~\cite{Liu:2007fg,Chen:2015kpa}:
\begin{eqnarray}
&& J^\alpha_{3/2,-,\Omega_b^-,2,1,\lambda}
\\ \nonumber &=& i \epsilon_{abc} \Big ( [\mathcal{D}_t^{\mu} s^{aT}] C \gamma_t^\nu s^b + s^{aT} C \gamma_t^\nu [\mathcal{D}_t^{\mu} s^b] \Big )
\\ \nonumber && ~~ \times \Big ( g_t^{\alpha\mu} \gamma_t^{\nu} \gamma_5 + g_t^{\alpha\nu} \gamma_t^{\mu} \gamma_5 - {2 \over 3} g_t^{\mu\nu} \gamma_t^{\alpha} \gamma_5 \Big ) h_v^c \, ,
\\ && J_{\Xi_b^{0}} = \epsilon_{abc} [d^{aT} C\gamma_{5} s^{b}] h_{v}^{c} \, ,
\end{eqnarray}
which couple to $\Omega_b^-({3/2}^-)$ and $\Xi_b^0(1/2^+)$, respectively.

\begin{widetext}
At the hadron level, we write $G^\alpha_{\Omega_b^-[{3\over2}^-] \rightarrow \Xi_b^{0}K^-}$ as:
\begin{eqnarray}
G^\alpha_{\Omega_b^-[{3\over2}^-] \rightarrow \Xi_b^{0}K^-} (\omega, \omega^\prime) &=& g_{\Omega_b^-[{3\over2}^-] \rightarrow \Xi_b^{0}K^-} \times { f_{\Omega_b^-[{3\over2}^-]} f_{\Xi_b^{0}} \over (\bar \Lambda_{\Omega_b^-[{3\over2}^-]} - \omega^\prime) (\bar \Lambda_{\Xi_b^{0}} - \omega)} \times \gamma \cdot q~\gamma_5~q^\alpha + \cdots \, , \label{G0C}
\end{eqnarray}
where $\cdots$ denotes other possible decay amplitudes.

At the quark-gluon level, we calculate $G^\alpha_{\Omega_b^-[{3\over2}^-] \rightarrow \Xi_b^{0}K^-}$ using the method of operator product expansion (OPE):
\begin{eqnarray}
\label{eq:sumrule}
&& G^\alpha_{\Omega_b^-[{3\over2}^-] \rightarrow \Xi_b^{0}K^-} (\omega, \omega^\prime)
\\ \nonumber &=& \int_0^\infty dt \int_0^1 du e^{i (1-u) \omega^\prime t} e^{i u \omega t} \times 8 \times \Big (
\frac{f_K m_s u}{4\pi^2 t^2}\phi_{2;K}(u)+\frac{f_K m_s^2 u}{12(m_u+m_s)\pi^2 t^2}\phi_{3;K}^\sigma(u)
\\ \nonumber &&
+ \frac{f_K m_s^2 m_K^2 u}{48(m_u+m_s)\pi^2}\phi_{3;K}^\sigma(u)+\frac{f_K m_s u}{64\pi^2}\phi_{4;K}(u)+\frac{f_K u}{12}\langle \bar s s\rangle\phi_{2;K}(u)+\frac{f_K m_s m_K^2 u t^2}{288(m_u+m_s)}\langle s s\rangle\phi_{3;K}^\sigma(u)
\\ \nonumber &&
+ \frac{f_K u t^2}{192}\langle s s\rangle\phi_{4;K}(u)+\frac{f_K u t^2}{192}\langle g_s \bar s\sigma G s\rangle \phi_{2;K}(u)+\frac{f_K u t^4}{3072}\langle g_s \bar s \sigma G s\rangle\phi_{4;K}(u) \Big ) \times \gamma \cdot q~\gamma_5~q^\alpha
\\ \nonumber &-&
\int_0^\infty dt \int_0^1 du \int \mathcal{D} \underline{\alpha} e^{i \omega^{\prime} t(\alpha_2 + u \alpha_3)} e^{i \omega t(1 - \alpha_2 - u \alpha_3)} \times \Big (\frac{f_{3K} u}{2\pi^2 t^2}\Phi_{3;K}(\underline{\alpha})-\frac{f_{3K}}{2\pi^2 t^2}\Phi_{3;K}(\underline{\alpha})
\\ \nonumber &&
+\frac{i f_{3K} u^2 \alpha_3}{2\pi^2 t v \cdot q}\Phi_{3;K}(\underline{\alpha})+\frac{i f_{3K} u \alpha_2}{2\pi^2 t v \cdot q}\Phi_{3;K}(\underline{\alpha})-\frac{i f_{3K} u}{2\pi^2 t v \cdot q}\Phi_{3;K}(\underline{\alpha})\Big ) \times \gamma \cdot q~\gamma_5~q^\alpha + \cdots\, .
\end{eqnarray}

Then we perform double Borel transformations to both Eqs.~(\ref{G0C}) and (\ref{eq:sumrule}), and obtain:
\begin{eqnarray}
&& g_{\Omega_b^-[{3\over2}^-] \rightarrow \Xi_b^{0}K^-} f_{\Omega_b^-[{3\over2}^-]} f_{\Xi_b^{0}} e^{- {\bar \Lambda_{\Omega_b^-[{3\over2}^-]} \over T_1}} e^{ - {\bar \Lambda_{\Xi_b^{0}} \over T_2}}
\label{eq:621lambda}
\\ \nonumber &=& 8 \times \Big ( -\frac{i f_k m_s u_0}{4\pi^2}T^3 f_2({\omega_c \over T})\phi_{2;K}(u_0)-\frac{i f_K m_K^2 u_0}{12(m_u+m_s)\pi^2}T^3 f_2({\omega_c \over T})\phi_{3;K}^\sigma(u_0)+\frac{i f_K m_s u_0}{64\pi^2}T f_0({\omega_c \over T})\phi_{4;K}(u_0)
\\ \nonumber &&
+\frac{i f_K u_0}{12}\langle \bar s s\rangle T f_0({\omega_c \over T})\phi_{2;K}(u_0)-\frac{i f_K m_s u_0}{288(m_u+m_s)}\langle \bar s s\rangle {1\over T}\phi_{3;K}^\sigma(u_0)-\frac{i f_K u_0}{192}\langle \bar s s\rangle {1\over T}\phi_{4;K}(u_0)
\\ \nonumber &&
-\frac{i f_K u_0}{192}\langle g_s \bar s\sigma G s\rangle {1\over T}\phi_{2;K}(u_0)+\frac{i f_K u_0}{3072}\langle g_s \bar s \sigma G s\rangle{1\over T^3}\phi_{4;K}(u_0) \Big )
\\ \nonumber &-&
\Big(-\frac{i f_{3K}}{2\pi^2}T^3f_2({\omega_c \over T}) \int_0^{1 \over 2} d\alpha_2 \int_{{1 \over 2}-\alpha_2}^{1-\alpha_2} d\alpha_3 ({u_0 \over \alpha_3} \Phi_{3;K}(\underline{\alpha})-{1 \over \alpha_3} \Phi_{3;K}(\underline{\alpha}))
\\ \nonumber &&
+\frac{i f_{3K}}{2\pi^2}T^3f_2({\omega_c\over T}) \int_0^{1 \over 2} d\alpha_2 \int_{{1 \over 2}-\alpha_2}^{1-\alpha_2} d\alpha_3 {1\over \alpha_3}{\partial\over\partial\alpha_3}(\alpha_3 u_0\Phi_{3;K}(\underline{\alpha})+\alpha_2\Phi_{3;K}(\underline{\alpha})-\Phi_{3;K}(\underline{\alpha}))\Big ) \, .
\end{eqnarray}
\end{widetext}
In the above expression, $\omega$ and $\omega^\prime$ have been transformed to $T_1$ and $T_2$; we work at the symmetric point $T_1 = T_2 = 2T$ so that $u_0 = {T_1 \over T_1 + T_2} = {1\over2}$; $f_n(x) \equiv 1 - e^{-x} \sum_{k=0}^n {x^k \over k!}$. We refer to Refs.~\cite{Ball:1998je,Ball:2006wn,Ball:2004rg,Ball:1998kk,Ball:1998sk,Ball:1998ff,Ball:2007rt,Ball:2007zt} for explicit forms of the light-cone distribution amplitudes contained in the above sum rule equations, and more examples can be found in Appendix~\ref{sec:othersumrule}.

In the present study we use the following values for various quark and gluon parameters at the renormalization scale 2~GeV~\cite{pdg,Yang:1993bp,Hwang:1994vp,Ovchinnikov:1988gk,Jamin:2002ev,Ioffe:2002be,Narison:2002pw,Gimenez:2005nt,colangelo}:
%
\begin{eqnarray}
\nonumber && \langle \bar qq \rangle = - (0.24 \pm 0.01 \mbox{ GeV})^3 \, ,
\\ \nonumber && \langle \bar ss \rangle = (0.8\pm 0.1)\times \langle\bar qq \rangle \, ,
\\ && \langle g_s \bar q \sigma G q \rangle = M_0^2 \times \langle \bar qq \rangle\, ,
\label{eq:condensates}
\\ \nonumber && \langle g_s \bar s \sigma G s \rangle = M_0^2 \times \langle \bar ss \rangle\, ,
\\ \nonumber && M_0^2= 0.8 \mbox{ GeV}^2\, ,
\\ \nonumber && \langle g_s^2GG\rangle =(0.48\pm 0.14) \mbox{ GeV}^4\, .
\end{eqnarray}
After fixing $\omega_c = 1.665$~GeV to be the average of the threshold values of the $\Omega_b^-({3/2}^-)$ and $\Xi_b^{0}$ mass sum rules, we calculate the coupling constant $g_{\Omega_b^-[{3\over2}^-] \rightarrow \Xi_b^{0}K^-}$ from Eq.~(\ref{eq:621lambda}) to be:
\begin{eqnarray}
\nonumber g_{\Omega_b^-[{3\over2}^-] \rightarrow \Xi_b^{0}K^-} &=& 7.27~{^{+1.10}_{-0.68}}~{^{+2.22}_{-1.67}}~{^{+2.34}_{-1.66}}~{^{+1.96}_{-1.42}}~{\rm GeV}^{-2}
\\ &=& 7.27~{^{+3.92}_{-2.83}}~{\rm GeV}^{-2} \, .
\end{eqnarray}
Here the uncertainties are due to the Borel mass, the parameters of $\Xi_b^{0}$, the parameters of $\Omega_b^-({3/2}^-)$, and various quark and gluon parameters listed in Eq.~(\ref{eq:condensates}), respectively. Some of these  parameters can be found in Sec.~\ref{sec:sumrule}. The variation of $g_{\Omega_b^-[{3\over2}^-] \rightarrow \Xi_b^{0}K^-}$ is shown in Fig.~\ref{fig:621lambda}(d) as a function of the Borel mass $T$, where we find that its Borel mass dependence is moderate and acceptable.

The $D$-wave decay of $\Omega_b^-({3/2}^-)$ into $\Xi_b^{0}K^-$ is kinematically allowed, and its amplitude is
\begin{eqnarray}
&& \mathcal{M} \left( \Omega_b^-({3/2}^-) \rightarrow \Xi_b^{0} + K^- \right)
\\ \nonumber &=& g_{\Omega_b^-[{3\over2}^-] \rightarrow \Xi_b^{0}K^-}~\bar u_{0}^{\mu} \gamma^\nu \gamma_5 u_1~p_{2,\mu} p_{2,\nu} \, ,
\end{eqnarray}
This amplitude can be used to further calculate its width through
\begin{eqnarray}
&& \Gamma \left( \Omega_b^-({3/2}^-) \rightarrow \Xi_b^{0} + K^- \right)
\\ \nonumber &=& \frac{|\vec p_2|}{32\pi^2 m_0^2} \times g_{\Omega_b^-[{3\over2}^-] \rightarrow \Xi_b^{0}K^-}^2 \times p_{2,\mu}p_{2,\nu}p_{2,\rho}p_{2,\sigma}
\\ \nonumber && \times ~{\rm Tr}\Big[ \gamma^\nu\gamma_5 \left( p\!\!\!\slash_1 + m_1 \right) \gamma^\sigma \gamma_5
\\ \nonumber && \Big(g^{\rho\mu}-{\gamma^\rho\gamma^\mu\over3}-{p_{0}^{\rho}\gamma^\mu-p_{0}^{\mu}\gamma^\rho \over 3m_0} -{2p_{0}^{\rho}p_{0}^{\mu} \over 3m_0^2}  \Big) ( p\!\!\!\slash_0 + m_0 ) \Big],
\end{eqnarray}
from which we obtain
\begin{eqnarray}
\Gamma_{\Omega_b^-[{3\over2}^-] \rightarrow \Xi_b^{0}K^-}&=& 4.6~{^{+3.3}_{-1.9}}{\rm~MeV} \, .
\end{eqnarray}

In the following subsections we shall follow the same procedures to separately investigate the four bottom baryon multiplets $[\mathbf{6}_F, 1, 0, \rho]$, $[\mathbf{6}_F, 0, 1, \lambda]$, $[\mathbf{6}_F, 1, 1, \lambda]$, and $[\mathbf{6}_F, 2, 1, \lambda]$.

\subsection{The $[\mathbf{6}_F, 1, 0, \rho]$ doublet}

\begin{figure*}[htb]
\begin{center}
\subfigure[]{
\scalebox{0.5}{\includegraphics{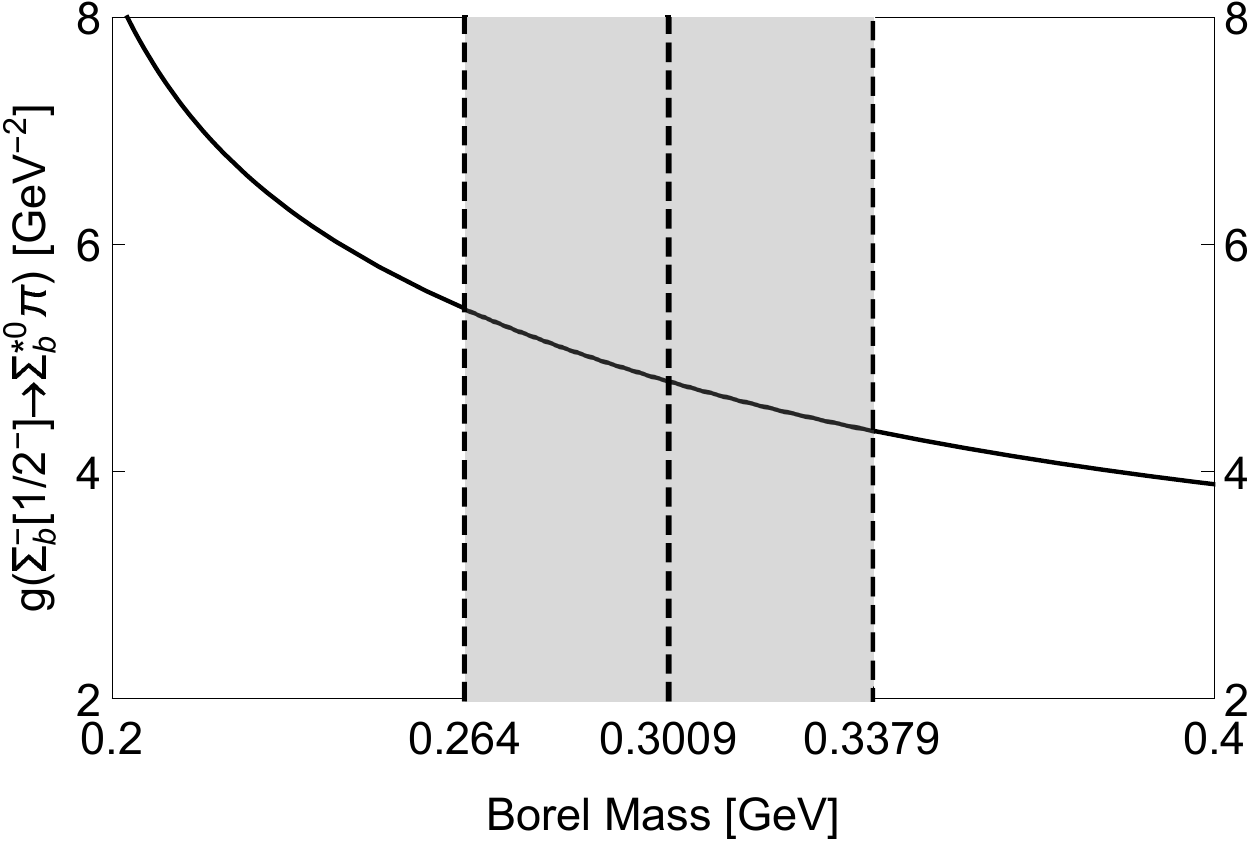}}}~~~~~
\subfigure[]{
\scalebox{0.5}{\includegraphics{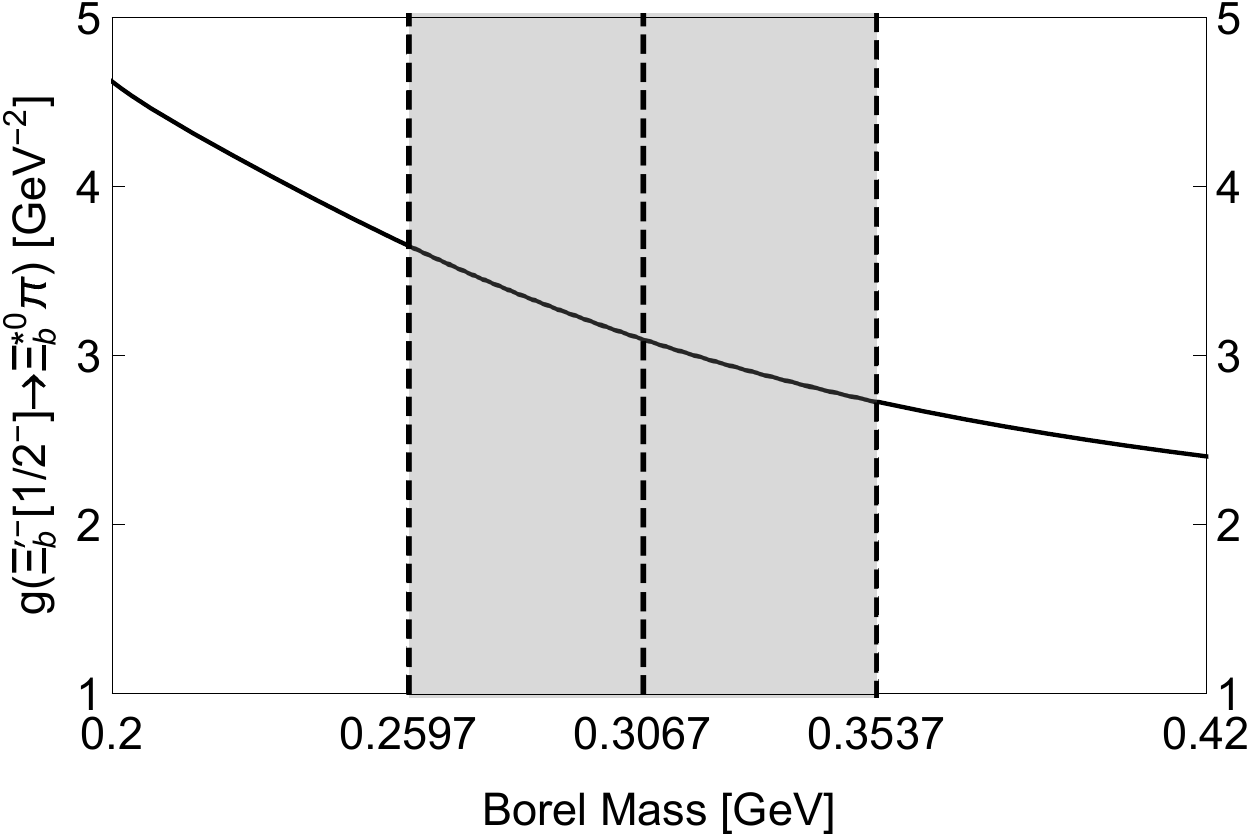}}}
\\
\subfigure[]{
\scalebox{0.5}{\includegraphics{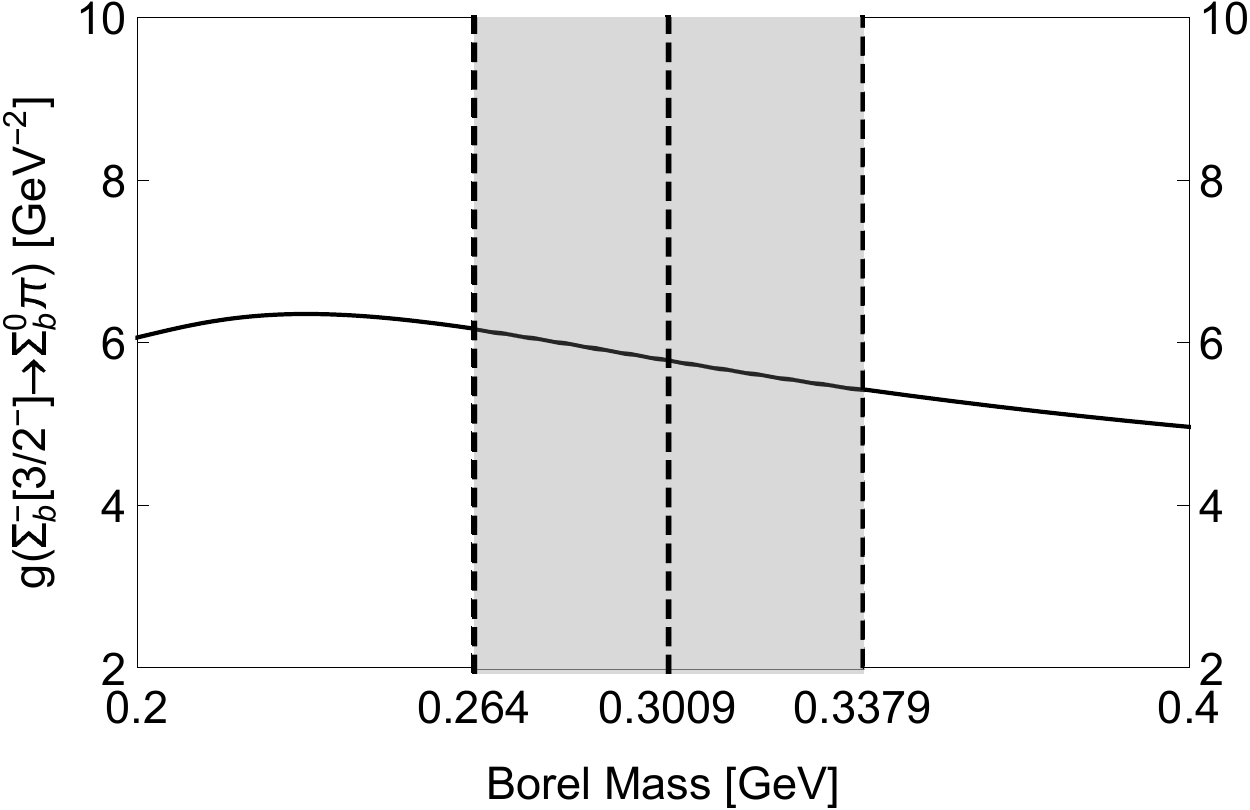}}}~~~~~
\subfigure[]{
\scalebox{0.5}{\includegraphics{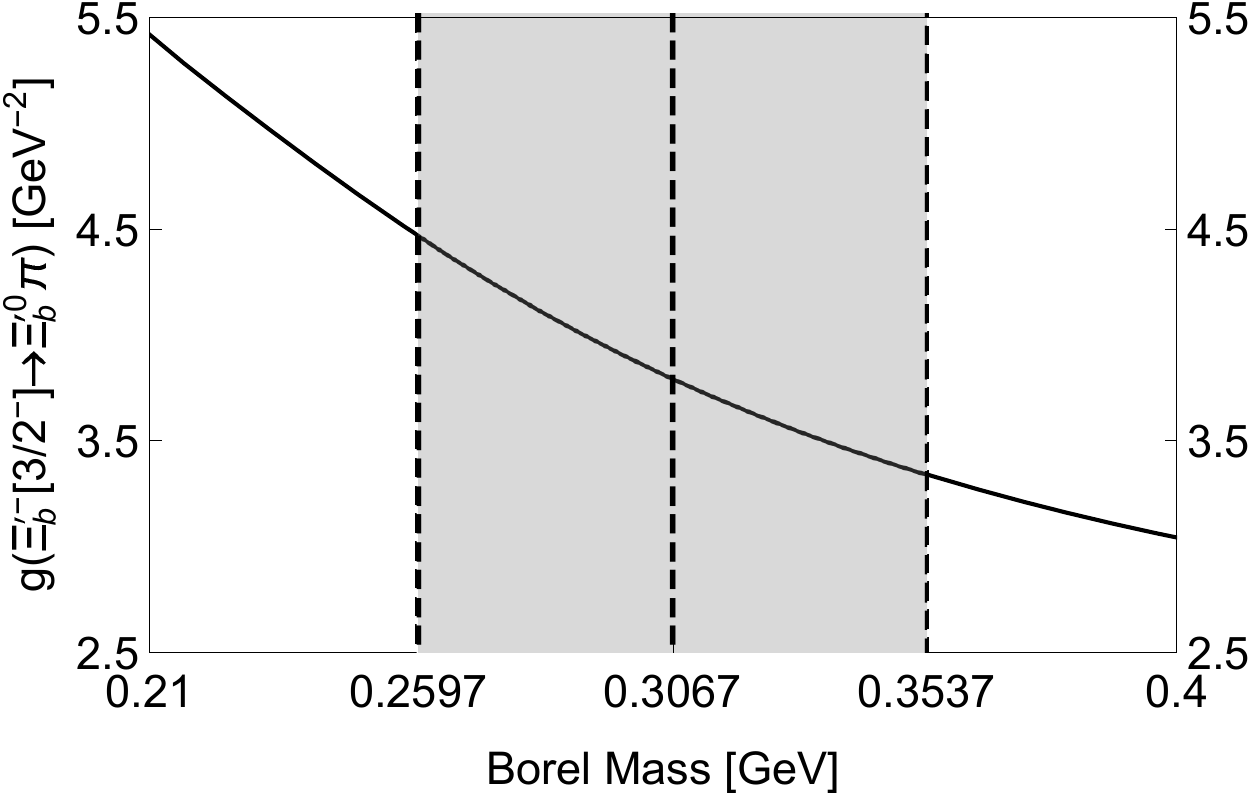}}}
\\
\subfigure[]{
\scalebox{0.5}{\includegraphics{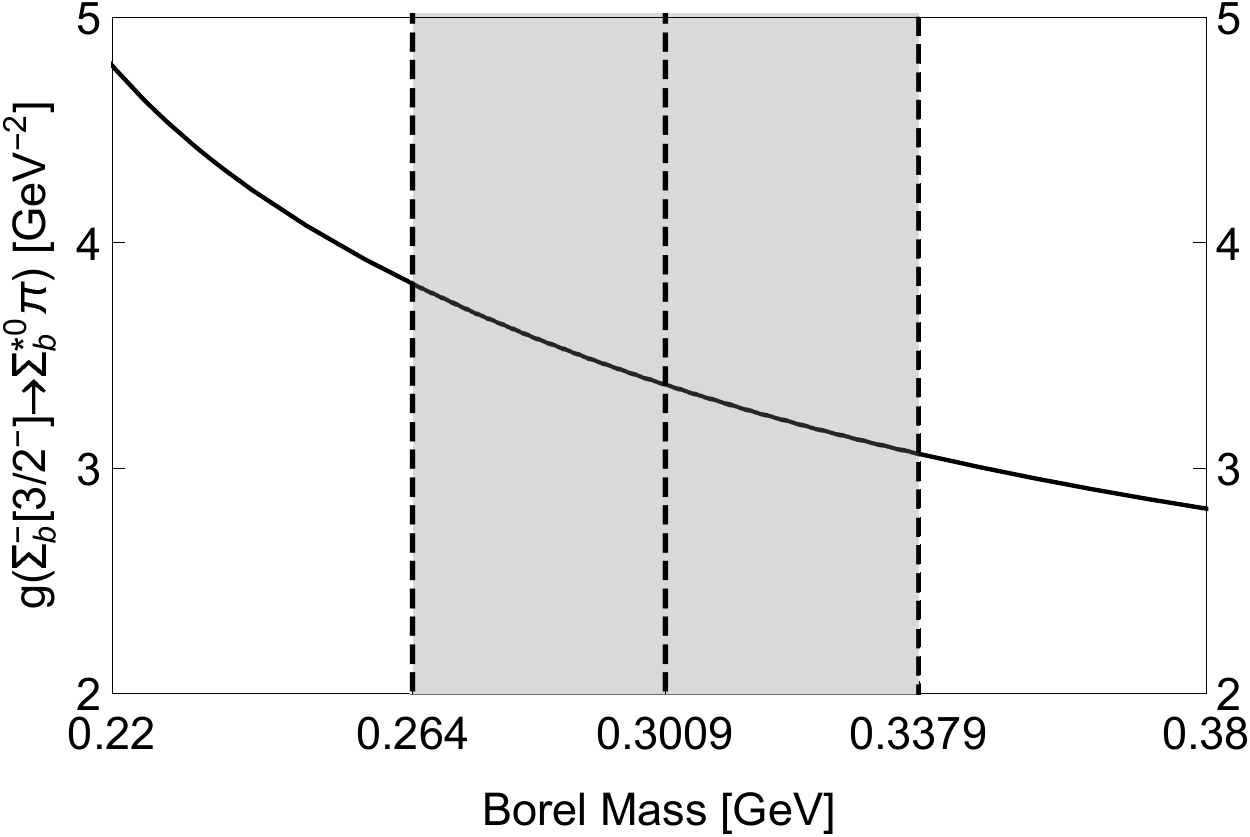}}}~~~~~
\subfigure[]{
\scalebox{0.5}{\includegraphics{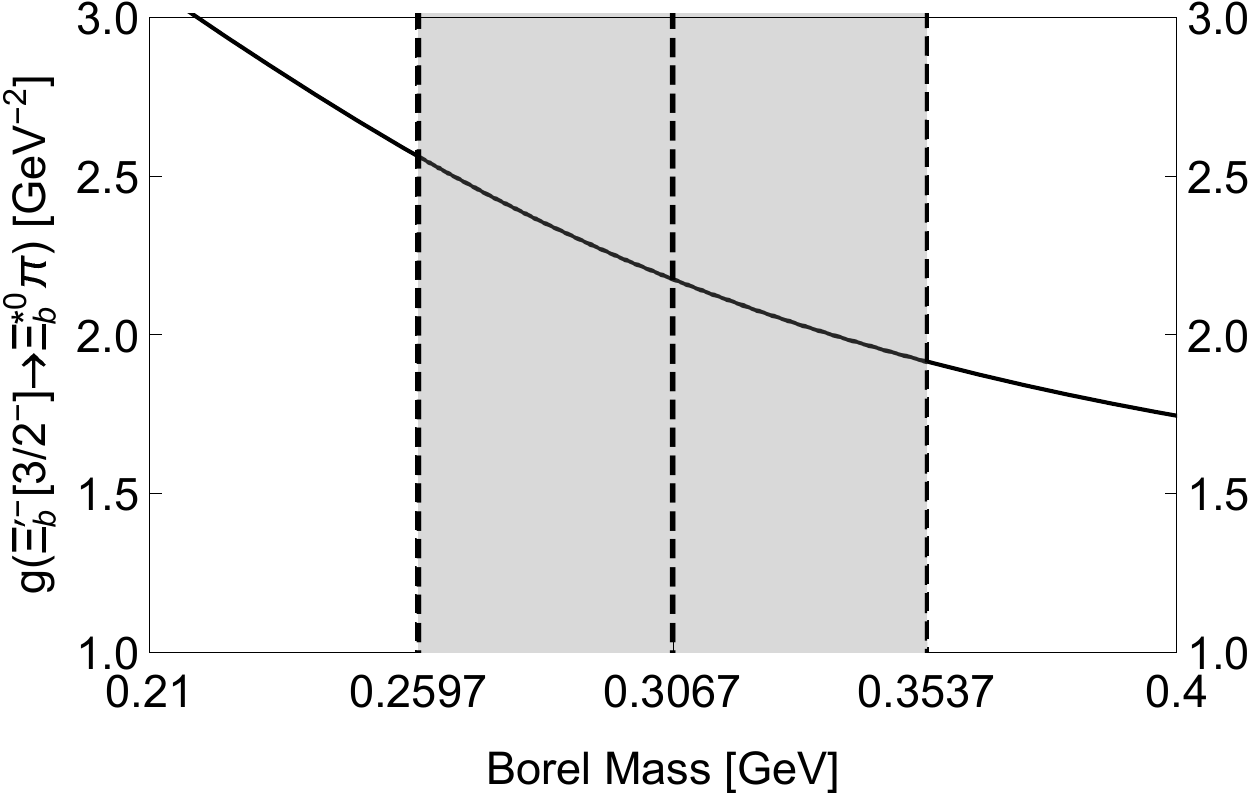}}}
\end{center}
\caption{The coupling constants as functions of the Borel mass $T$: (a) $g_{\Sigma_b^-[{1\over2}^-] \rightarrow \Sigma_b^{*0}\pi^-}$, (b) $g_{\Xi_b^{\prime-}[{1\over2}^-] \rightarrow \Xi_b^{*0} \pi^-}$, (c) $g_{\Sigma_b^-[{3\over2}^-] \rightarrow \Sigma_b^0 \pi^-}$, (d) $g_{\Xi_b^{\prime-}[{3\over2}^-] \rightarrow \Xi_b^{\prime0} \pi^-}$, (e) $g_{\Sigma_b^-[{3\over2}^-] \rightarrow \Sigma_b^{*0} \pi^-}$, and (f) $g_{\Xi_b^{\prime-}[{3\over2}^-] \rightarrow \Xi_b^{*0} \pi^-}$. Here the bottom baryon doublet $[\mathbf{6}_F, 1, 0, \rho]$ is investigated.
\label{fig:610rho}}
\end{figure*}

There are six bottom baryons contained in the $[\mathbf{6}_F, 1, 0, \rho]$ doublet, that are $\Sigma_b({1\over2}^-/{3\over2}^-)$, $\Xi^\prime_b({1\over2}^-/{3\over2}^-)$, and $\Omega_b({1\over2}^-/{3\over2}^-)$. We study their $D$-wave decays into ground-state bottom baryons and pseudoscalar mesons, and find twelve non-zero coupling constants:
\begin{eqnarray}
\nonumber &(a1)& g_{\Sigma_b[{1\over2}^-]\to \Sigma_b^{*}[{3\over2}^+] \pi} = 4.80~{\rm GeV}^{-2} \, ,
\\
\nonumber &(b2)& g_{\Sigma_b[{3\over2}^-]\to \Sigma_b[{1\over2}^+] \pi} = 5.79~{\rm GeV}^{-2}\, ,
\\
\nonumber &(b3)& g_{\Sigma_b[{3\over2}^-]\to \Sigma_b^{*}[{3\over2}^+] \pi}= 3.37~{\rm GeV}^{-2} \, ,
\\
\nonumber &(d1)& g_{\Xi_b^{\prime}[{1\over2}^-]\to \Xi_b^{*}[{3\over2}^+] \pi} = 3.10~{\rm GeV}^{-2} \, ,
\\
\nonumber &(d2)& g_{\Xi_b^{\prime}[{1\over2}^-]\to \Sigma_b^{*}[{3\over2}^+] K}= 3.63~{\rm GeV}^{-2} \, ,
\\
\nonumber &(e3)& g_{\Xi_b^{\prime}[{3\over2}^-]\to \Xi_b^{\prime}[{1\over2}^+] \pi } = 3.79~{\rm GeV}^{-2} \, ,
\\        &(e4)& g_{\Xi_b^{\prime}[{3\over2}^-]\to \Sigma_b[{1\over2}^+]K}= 4.45~{\rm GeV}^{-2} \, ,
\\
\nonumber &(e5)& g_{\Xi_b^{\prime}[{3\over2}^-]\to \Xi_b^{*}[{3\over2}^+]\pi}= 2.18~{\rm GeV}^{-2} \, ,
\\
\nonumber &(e6)& g_{\Xi_b^{\prime}[{3\over2}^-]\to\Sigma_b^{*}[{3\over2}^+]K}= 2.55~{\rm GeV}^{-2} \, ,
\\
\nonumber &(g1)& g_{\Omega_b[{1\over2}^-]\to \Xi_b^{*}[{3\over2}^+]K}= 4.54~{\rm GeV}^{-2} \, ,
\\
\nonumber &(h2)& g_{\Omega_b[{3\over2}^-]\to \Xi_b^{\prime}[{1\over2}^+]K}= 5.56~{\rm GeV}^{-2} \, ,
\\
\nonumber &(h3)& g_{\Omega_b[{3\over2}^-]\to \Xi_b^{*}[{3\over2}^+]K}= 3.19~{\rm GeV}^{-2} \, .
\end{eqnarray}
We show some of these coupling constants as functions of the Borel mass $T$ in Fig.~\ref{fig:610rho}. Based on them, we further find six $D$-wave decay channels that are kinematically allowed:
\begin{eqnarray}
\nonumber &(a1)& \Gamma_{\Sigma_b[{1\over2}^-]\to \Sigma_b^{*}[{3\over2}^+] \pi} =0.62~{\rm MeV} \, ,
\\
\nonumber &(b2)& \Gamma_{\Sigma_b[{3\over2}^-]\to \Sigma_b[{1\over2}^+] \pi} = 0.84~{\rm MeV} \, ,
\\
\nonumber &(b3)& \Gamma_{\Sigma_b[{3\over2}^-]\to \Sigma_b^{*}[{3\over2}^+] \pi} = 0.098~{\rm MeV} \, ,
\\
          &(d1)& \Gamma_{\Xi_b^{\prime}[{1\over2}^-]\to \Xi_b^{*}[{3\over2}^+] \pi}=0.29~{\rm MeV} \, ,
\\
\nonumber &(e3)& \Gamma_{\Xi_b^{\prime}[{3\over2}^-]\to \Xi_b^{\prime}[{1\over2}^+] \pi} = 0.47~{\rm MeV} \, ,
\\
\nonumber &(e5)& \Gamma_{\Xi_b^{\prime}[{3\over2}^-]\to \Xi_b^{*}[{3\over2}^+]\pi}=0.064~{\rm MeV } \, .
\end{eqnarray}
We summarize these $D$-wave decay widths in Table~\ref{tab:decay610rho}, where possible experimental candidates are given for comparisons. In Refs.~\cite{Chen:2015kpa,Mao:2015gya} we have studied the mass spectrum of $P$-wave bottom baryons, and the results are reanalysed and summarized in this table. In Refs.~\cite{Chen:2017sci,Yang:2019cvw} we have studied $S$-wave decay properties of $P$-wave bottom baryons into ground-state bottom baryons together with pseudoscalar mesons or vector mesons, and the results are also reanalysed and summarized in this table.

\subsection{The $[\mathbf{6}_F,0,1,\lambda]$ singlet}

There are three bottom baryons contained in the $[\mathbf{6}_F,0,1,\lambda]$ singlet, that are $\Sigma_b({1\over2}^-)$, $\Xi^\prime_b({1\over2}^-)$, and $\Omega_b({1\over2}^-)$. We study their $D$-wave decays into ground-state bottom baryons and pseudoscalar mesons, but find all the coupling constants to be zero. For completeness, we summarize these results in Table~\ref{tab:decay601lambda}, together with their mass spectrum, $S$-wave decay properties, and possible experimental candidates.

\subsection{The $[\mathbf{6}_F, 1, 1, \lambda]$ doublet}

\begin{figure*}[htb]
\begin{center}
\subfigure[]{
\scalebox{0.5}{\includegraphics{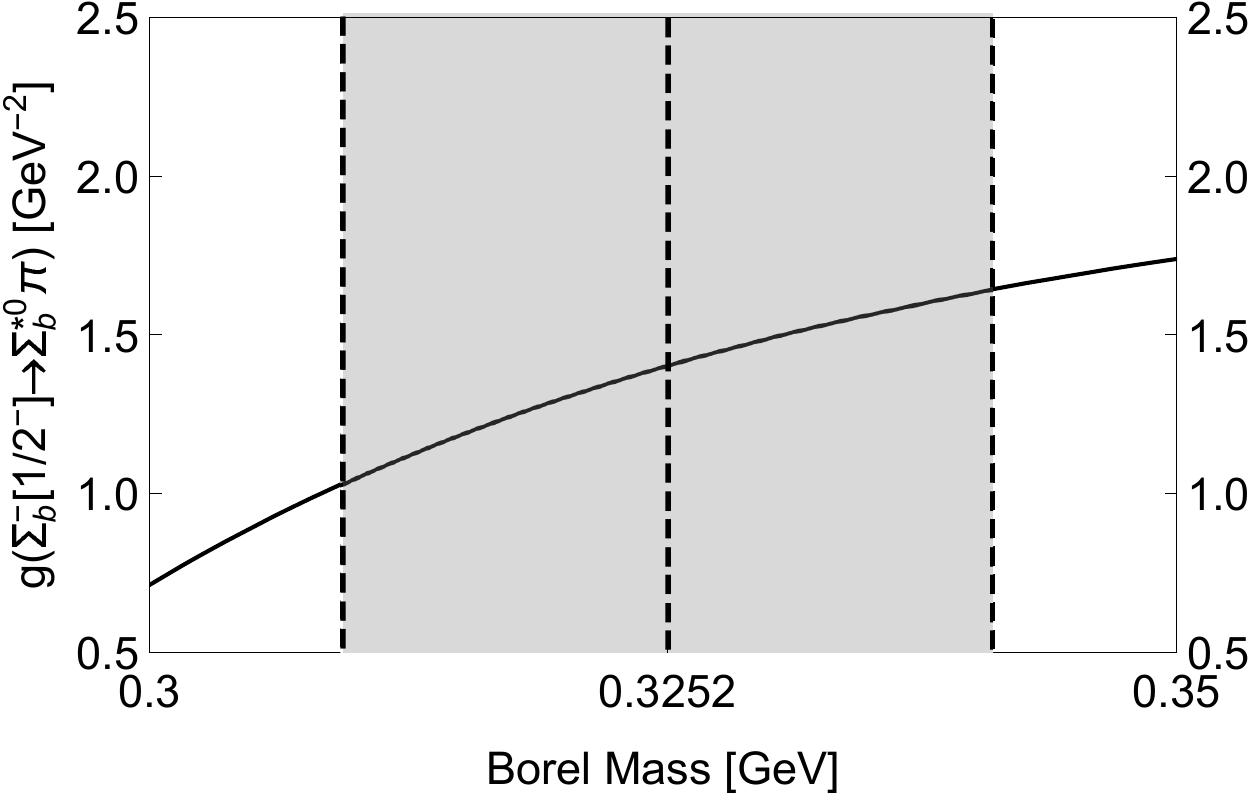}}}~~~~~
\subfigure[]{
\scalebox{0.5}{\includegraphics{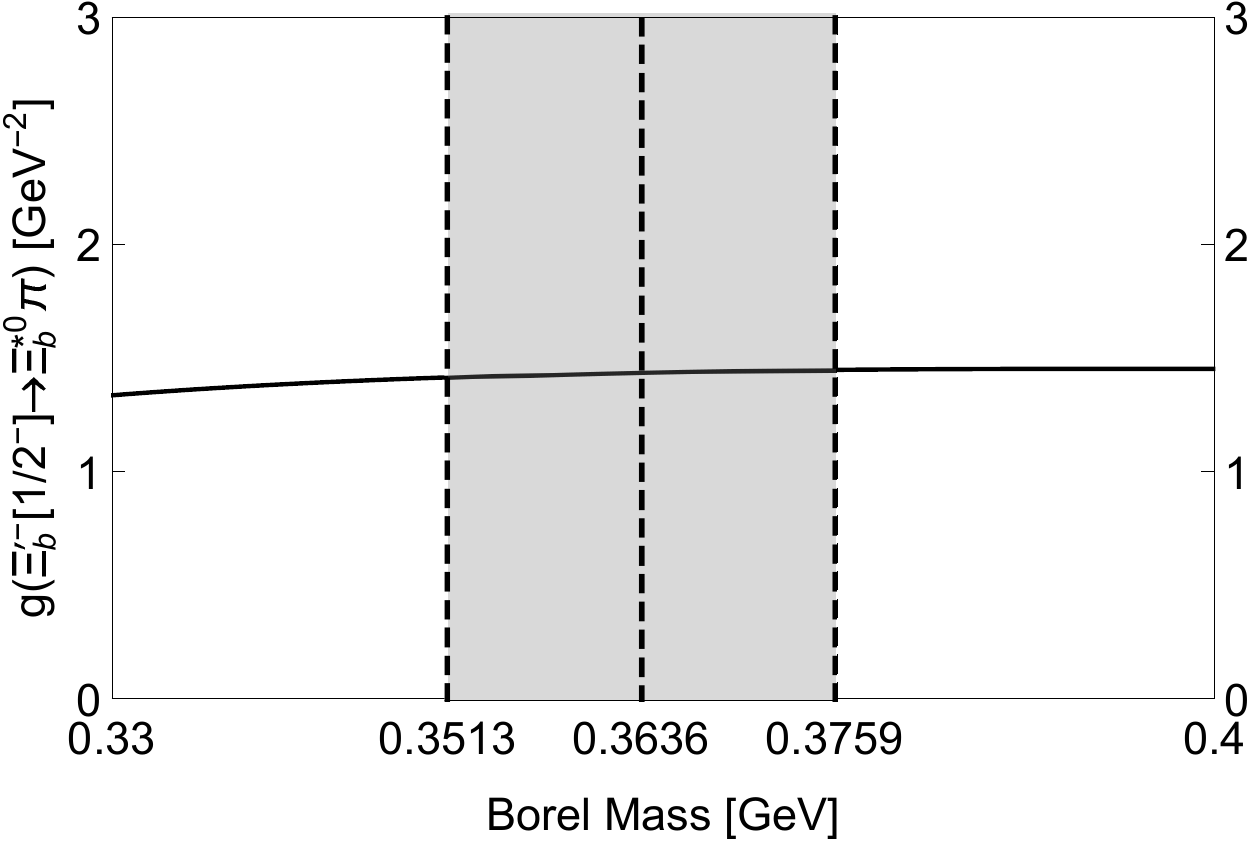}}}
\\
\subfigure[]{
\scalebox{0.5}{\includegraphics{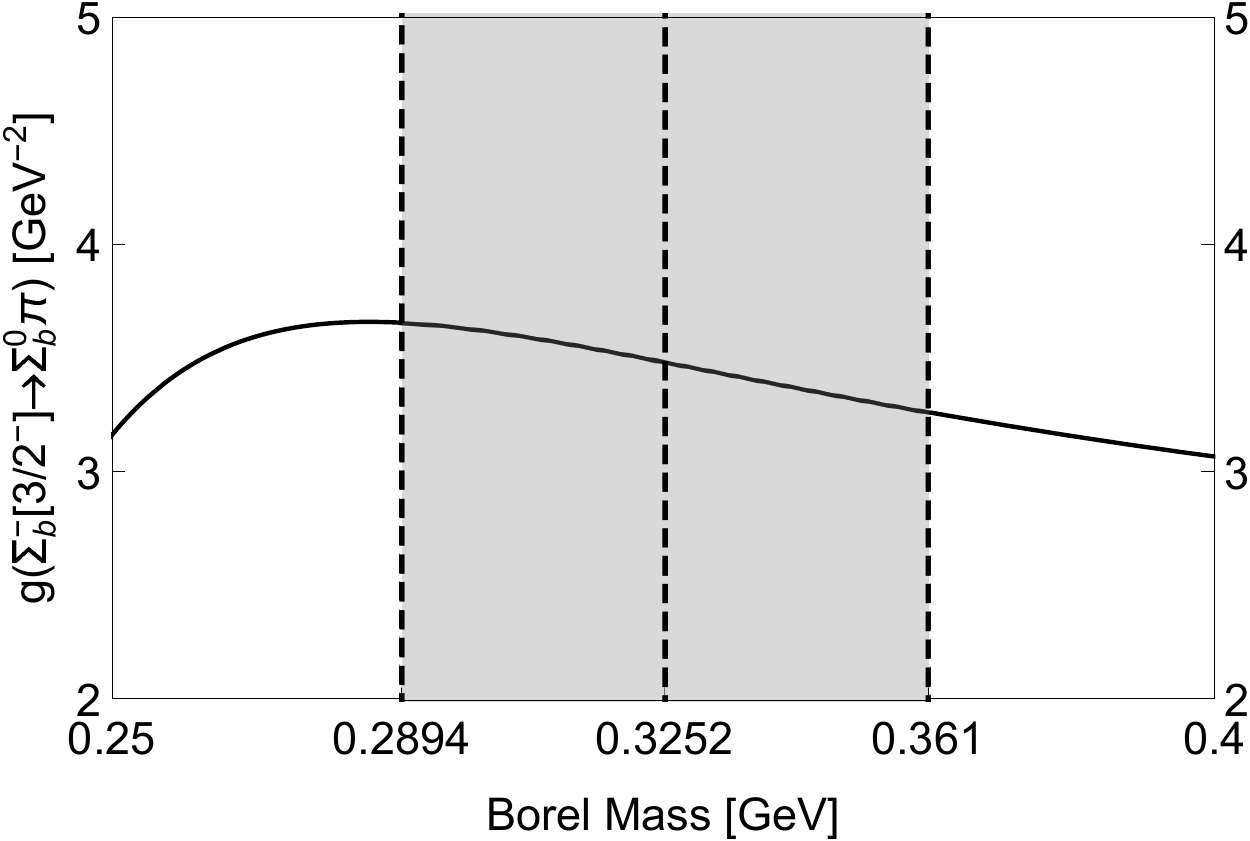}}}~~~~~
\subfigure[]{
\scalebox{0.5}{\includegraphics{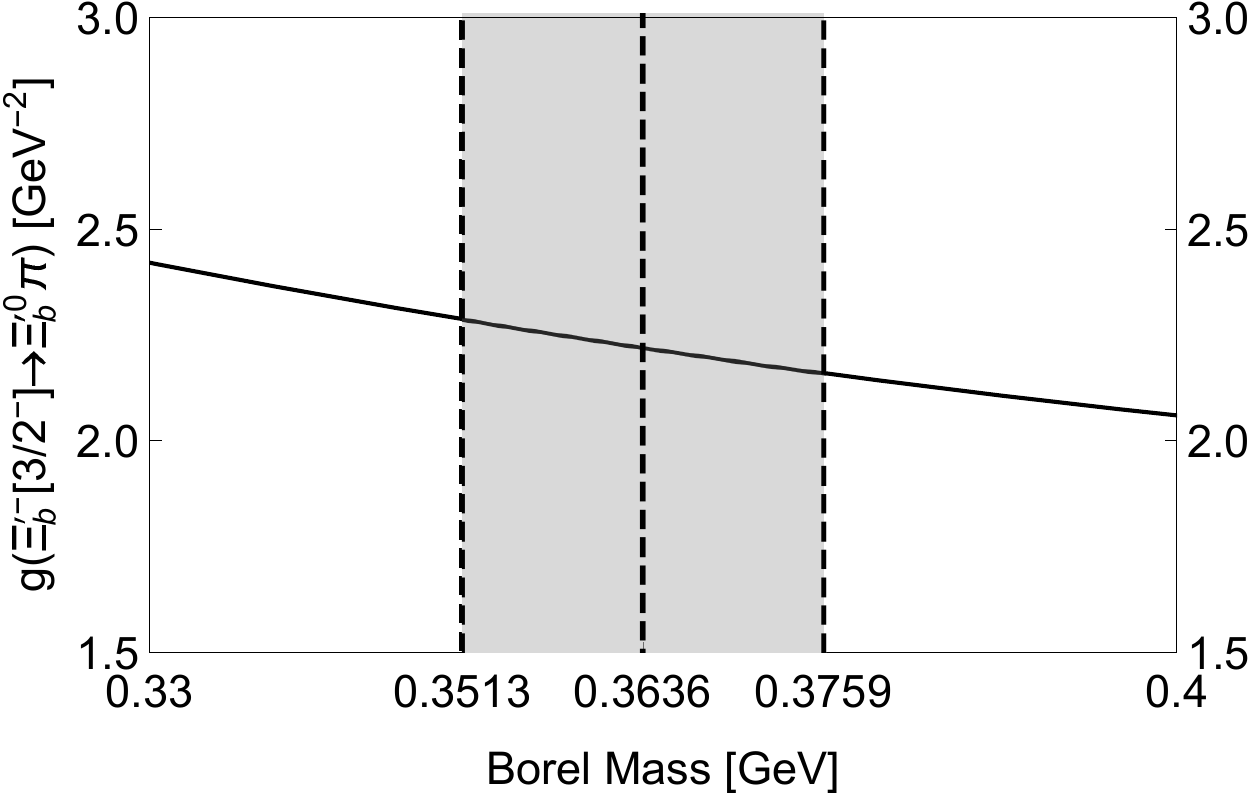}}}
\\
\subfigure[]{
\scalebox{0.5}{\includegraphics{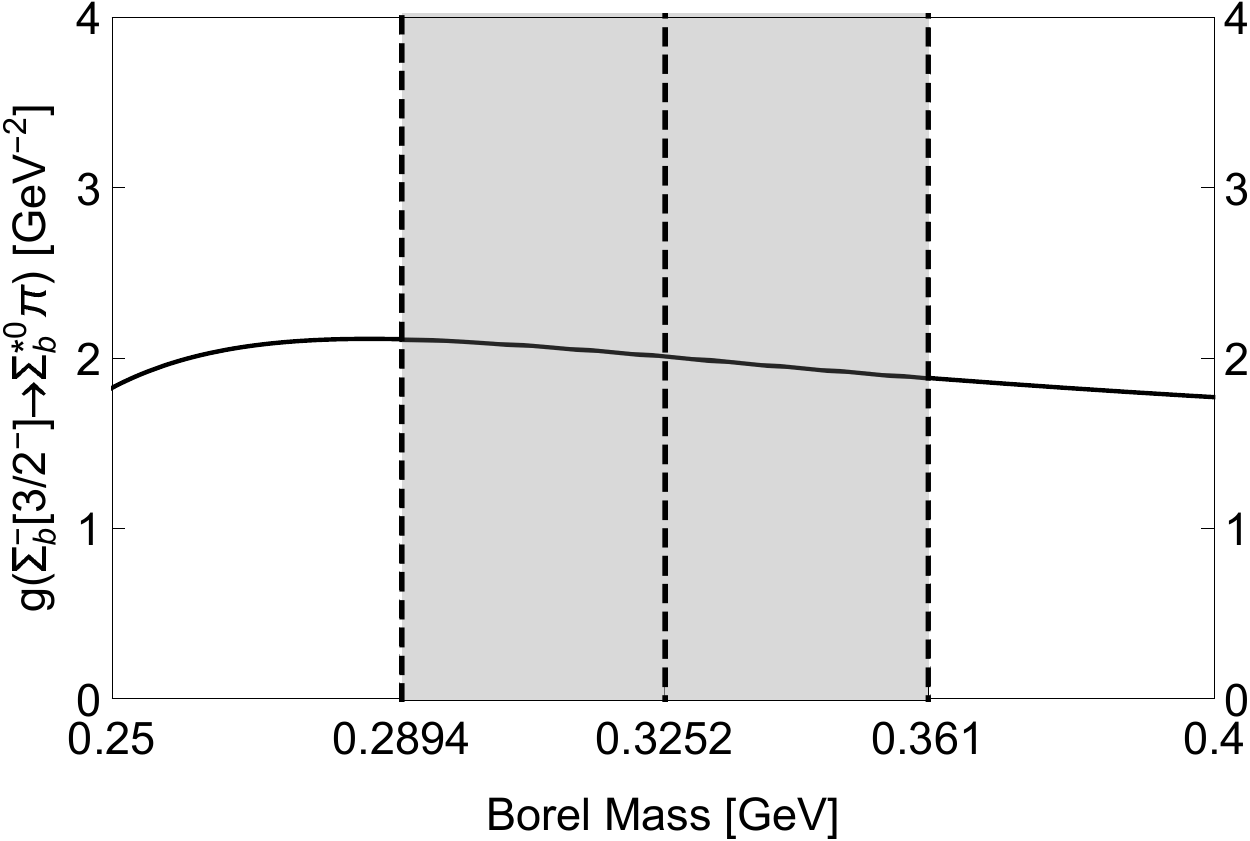}}}~~~~~
\subfigure[]{
\scalebox{0.5}{\includegraphics{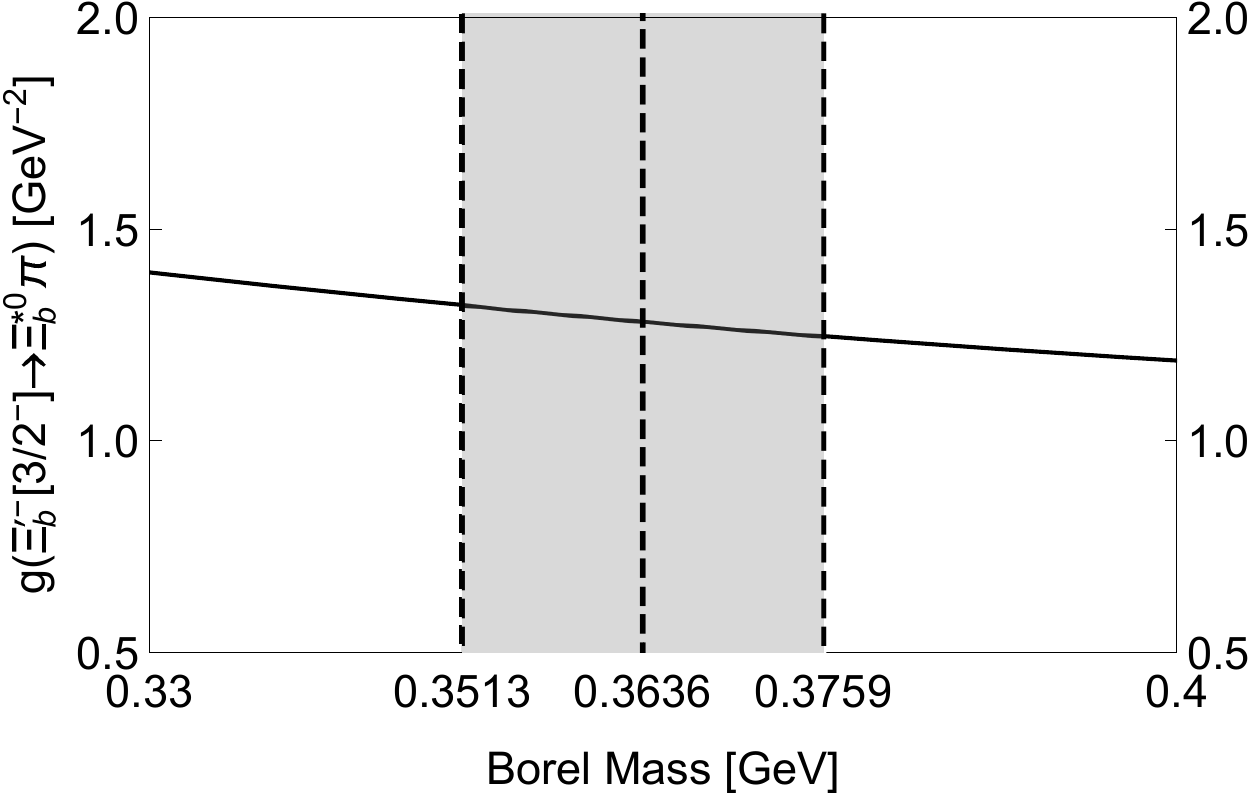}}}
\end{center}
\caption{The coupling constants as functions of the Borel mass $T$: (a) $g_{\Sigma_b^-[{1\over2}^-] \rightarrow \Sigma_b^{*0}\pi^-}$, (b) $g_{\Xi_b^{\prime-}[{1\over2}^-] \rightarrow \Xi_b^{*0} \pi^-}$, (c) $g_{\Sigma_b^-[{3\over2}^-] \rightarrow \Sigma_b^0 \pi^-}$, (d) $g_{\Xi_b^{\prime-}[{3\over2}^-] \rightarrow \Xi_b^{\prime0} \pi^-}$, (e) $g_{\Sigma_b^-[{3\over2}^-] \rightarrow \Sigma_b^{*0} \pi^-}$, and (f) $g_{\Xi_b^{\prime-}[{3\over2}^-] \rightarrow \Xi_b^{*0} \pi^-}$.
Here the bottom baryon doublet $[\mathbf{6}_F, 1, 1, \lambda]$ is investigated.
\label{fig:611lambda}}
\end{figure*}

There are six bottom baryons contained in the $[\mathbf{6}_F, 1, 1, \lambda]$ doublet, that are $\Sigma_b({1\over2}^-/{3\over2}^-)$, $\Xi^\prime_b({1\over2}^-/{3\over2}^-)$, and $\Omega_b({1\over2}^-/{3\over2}^-)$. We study their $D$-wave decays into ground-state bottom baryons and pseudoscalar mesons, and find twelve non-zero coupling constants:
\begin{eqnarray}
\nonumber &(a1)& g_{\Sigma_b[{1\over2}^-]\to \Sigma_b^{*}[{3\over2}^+] \pi} = 1.40~{^{+0.98}_{-1.40}}~{\rm GeV}^{-2} \, ,
\\
\nonumber &(b2)& g_{\Sigma_b[{3\over2}^-]\to \Sigma_b[{1\over2}^+] \pi} = 3.48~{^{+1.86}_{-1.47}}~{\rm GeV}^{-2}\, ,
\\
\nonumber &(b3)& g_{\Sigma_b[{3\over2}^-]\to \Sigma_b^{*}[{3\over2}^+] \pi}= 2.01~{^{+1.08}_{-0.85}}~{\rm GeV}^{-2} \, ,
\\
\nonumber &(d1)& g_{\Xi_b^{\prime}[{1\over2}^-]\to \Xi_b^{*}[{3\over2}^+] \pi} = 1.44~{^{+0.66}_{-0.61}}~{\rm GeV}^{-2} \, ,
\\
\nonumber &(d2)& g_{\Xi_b^{\prime}[{1\over2}^-]\to \Sigma_b^{*}[{3\over2}^+] K}= 0.78~{\rm GeV}^{-2} \, ,
\\
\nonumber &(e3)& g_{\Xi_b^{\prime}[{3\over2}^-]\to \Xi_b^{\prime}[{1\over2}^+] \pi } = 2.22~{^{+0.96}_{-0.80}}~{\rm GeV}^{-2} \, ,
\\        &(e4)& g_{\Xi_b^{\prime}[{3\over2}^-]\to \Sigma_b[{1\over2}^+]K}= 1.74~{\rm GeV}^{-2} \, ,
\\
\nonumber &(e5)& g_{\Xi_b^{\prime}[{3\over2}^-]\to \Xi_b^{*}[{3\over2}^+]\pi}= 1.28~{^{+0.58}_{-0.46}}~{\rm GeV}^{-2} \, ,
\\
\nonumber &(e6)& g_{\Xi_b^{\prime}[{3\over2}^-]\to\Sigma_b^{*}[{3\over2}^+]K}= 1.01~{\rm GeV}^{-2} \, ,
\\
\nonumber &(g1)& g_{\Omega_b[{1\over2}^-]\to \Xi_b^{*}[{3\over2}^+]K}= 1.59~{\rm GeV}^{-2} \, ,
\\
\nonumber &(h2)& g_{\Omega_b[{3\over2}^-]\to \Xi_b^{\prime}[{1\over2}^+]K}= 2.49~{\rm GeV}^{-2} \, ,
\\
\nonumber &(h3)& g_{\Omega_b[{3\over2}^-]\to \Xi_b^{*}[{3\over2}^+]K}= 1.44~{\rm GeV}^{-2} \, .
\end{eqnarray}
We show some of these coupling constants as functions of the Borel mass $T$ in Fig.~\ref{fig:611lambda}. Based on them, we further find six $D$-wave decay channels that are kinematically allowed:
\begin{eqnarray}
\nonumber &(a1)& \Gamma_{\Sigma_b[{1\over2}^-]\to \Sigma_b^{*}[{3\over2}^+] \pi} =0.076~{^{+0.144}_{-0.076}}~{\rm MeV} \, ,
\\
\nonumber &(b2)& \Gamma_{\Sigma_b[{3\over2}^-]\to \Sigma_b[{1\over2}^+] \pi} = 0.55~{^{+0.74}_{-0.36}}~{\rm MeV} \, ,
\\
\nonumber &(b3)& \Gamma_{\Sigma_b[{3\over2}^-]\to \Sigma_b^{*}[{3\over2}^+] \pi} = 0.070~{^{+0.096}_{-0.047}}~{\rm MeV} \, ,
\\
          &(d1)& \Gamma_{\Xi_b^{\prime}[{1\over2}^-]\to \Xi_b^{*}[{3\over2}^+] \pi}=0.16~{^{+0.18}_{-0.10}}~{\rm MeV} \, ,
\\
\nonumber &(e3)& \Gamma_{\Xi_b^{\prime}[{3\over2}^-]\to \Xi_b^{\prime}[{1\over2}^+] \pi} = 0.34~{^{+0.35}_{-0.20}}~{\rm MeV} \, ,
\\
\nonumber &(e5)& \Gamma_{\Xi_b^{\prime}[{3\over2}^-]\to \Xi_b^{*}[{3\over2}^+]\pi}= 0.051~{^{+0.057}_{-0.030}}~{\rm MeV } \, .
\end{eqnarray}
We summarize these $D$-wave decay widths in Table~\ref{tab:decay611lambda}, together with their mass spectrum, $S$-wave decay properties, and possible experimental candidates.

\subsection{The $[\mathbf{6}_F,2,1,\lambda]$ doublet}

\begin{figure*}[htb]
\begin{center}
\subfigure[]{
\scalebox{0.5}{\includegraphics{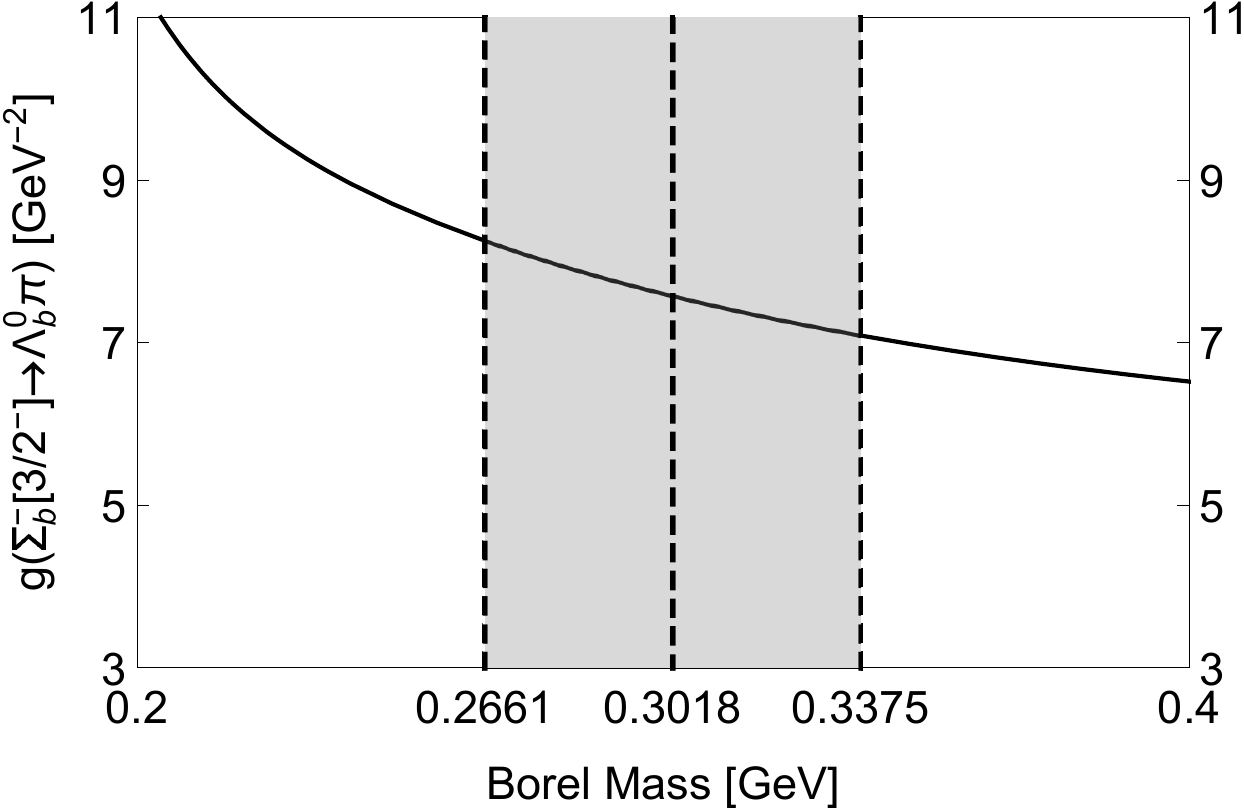}}}~~~~~
\subfigure[]{
\scalebox{0.5}{\includegraphics{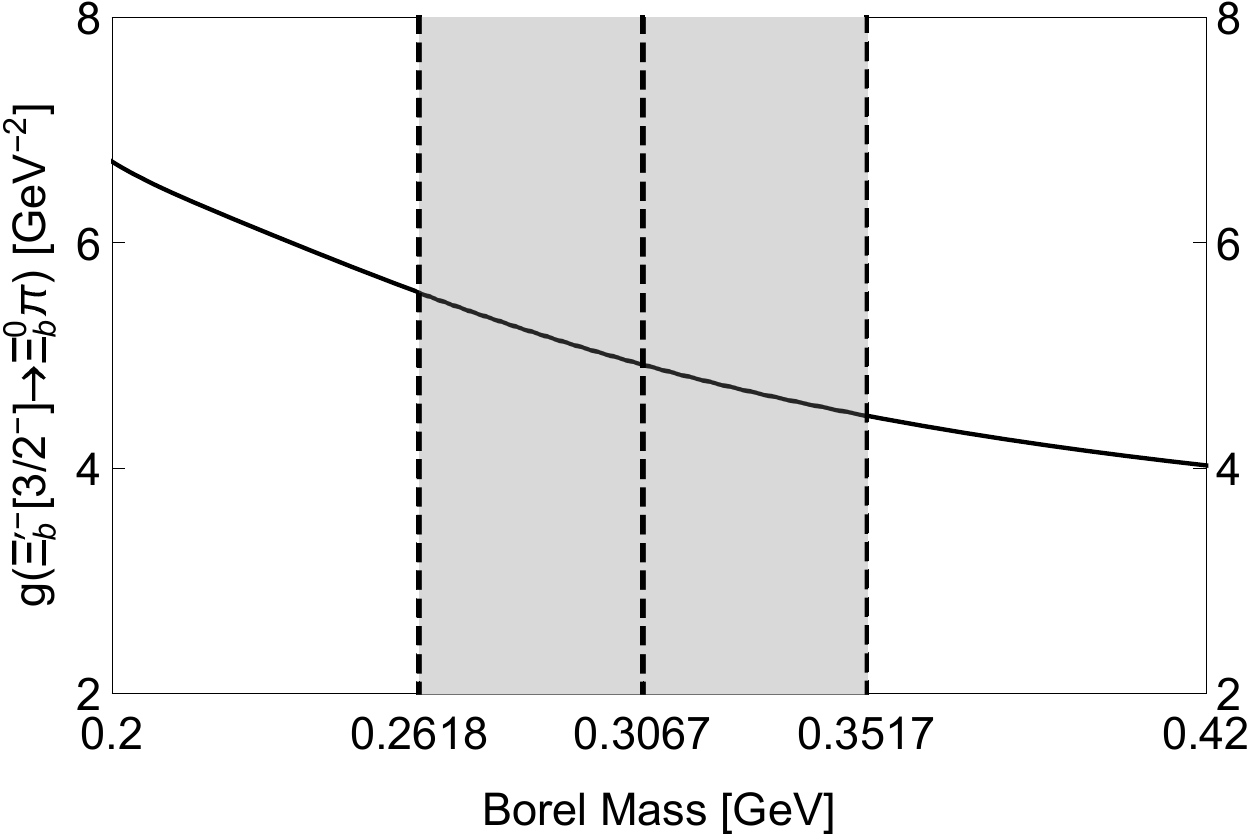}}}
\\
\subfigure[]{
\scalebox{0.5}{\includegraphics{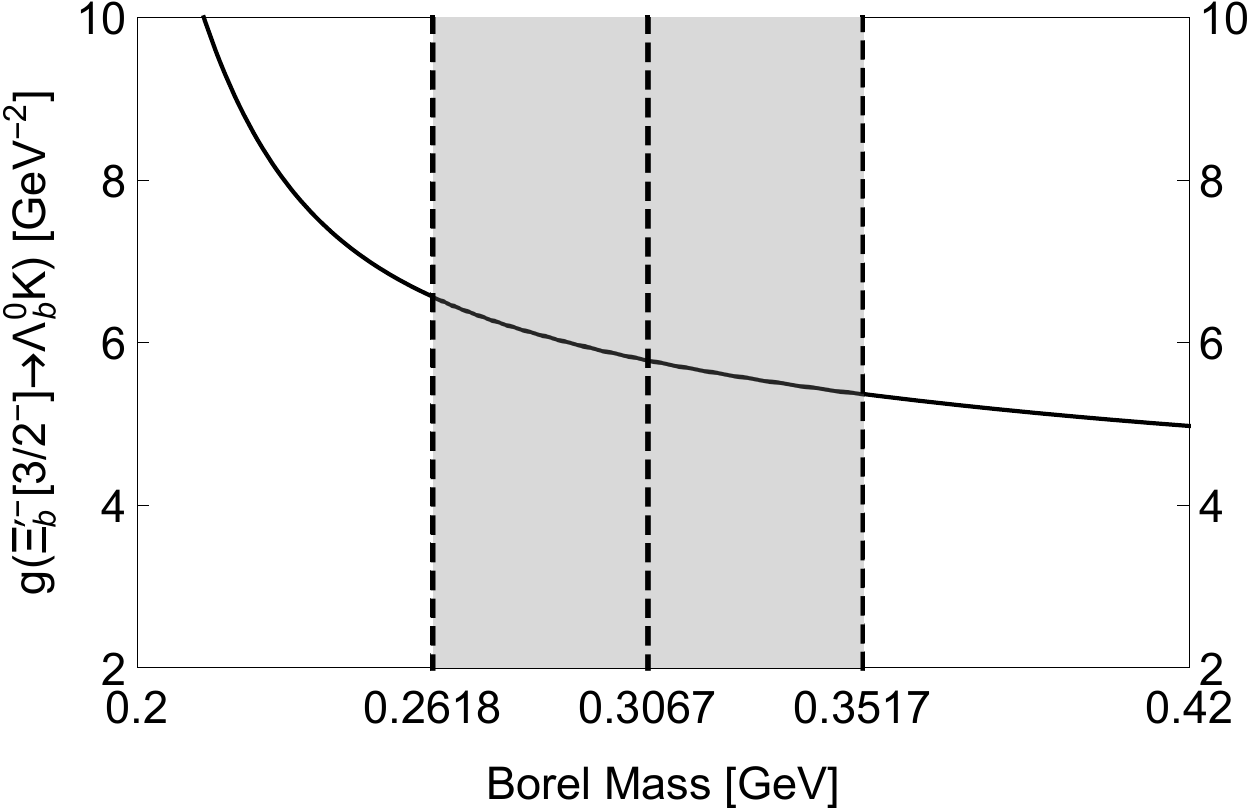}}}~~~~~
\subfigure[]{
\scalebox{0.5}{\includegraphics{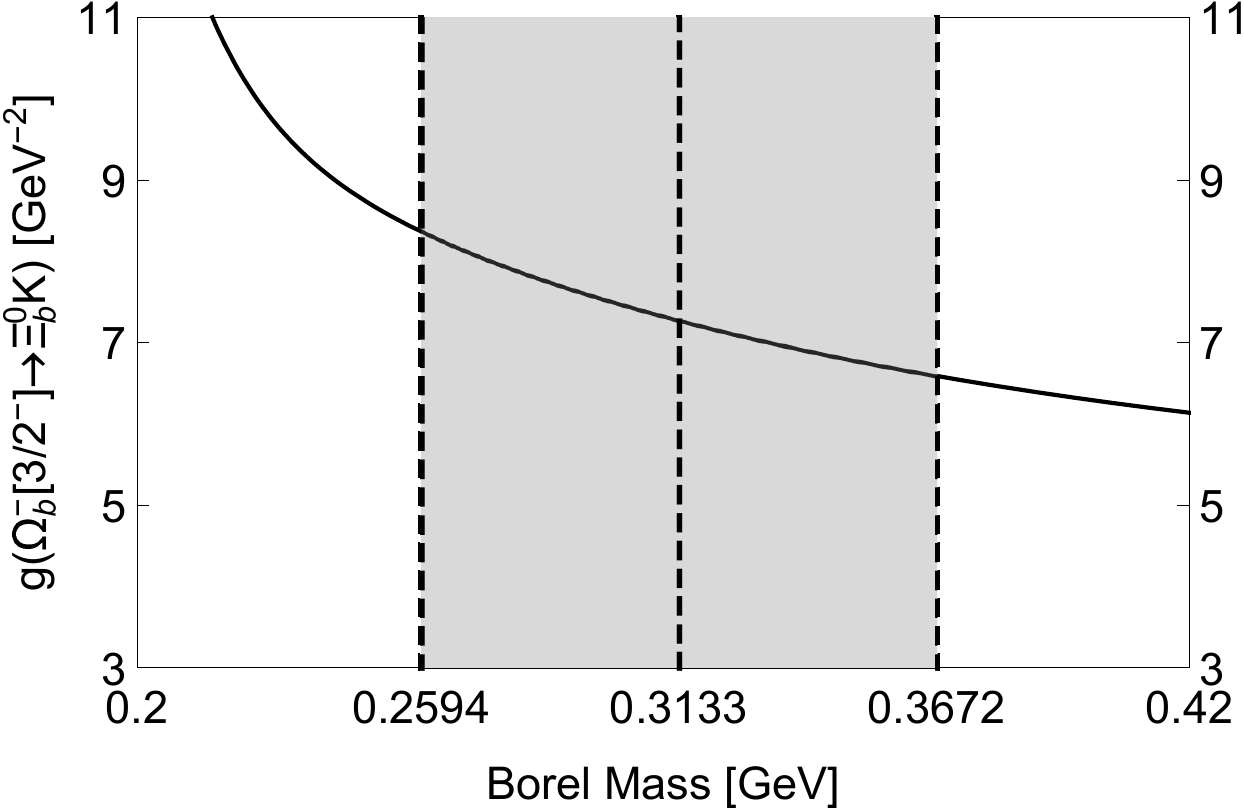}}}
\\
\subfigure[]{
\scalebox{0.5}{\includegraphics{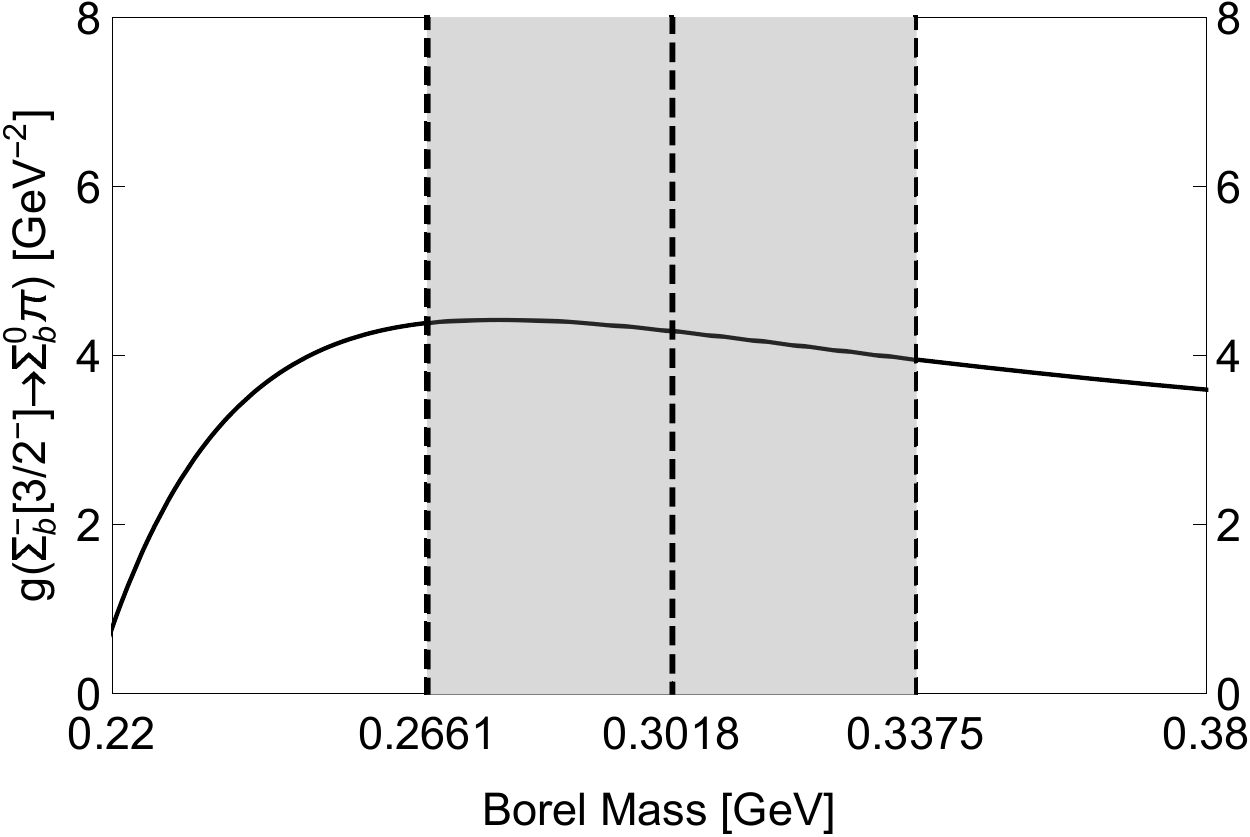}}}~~~~~
\subfigure[]{
\scalebox{0.5}{\includegraphics{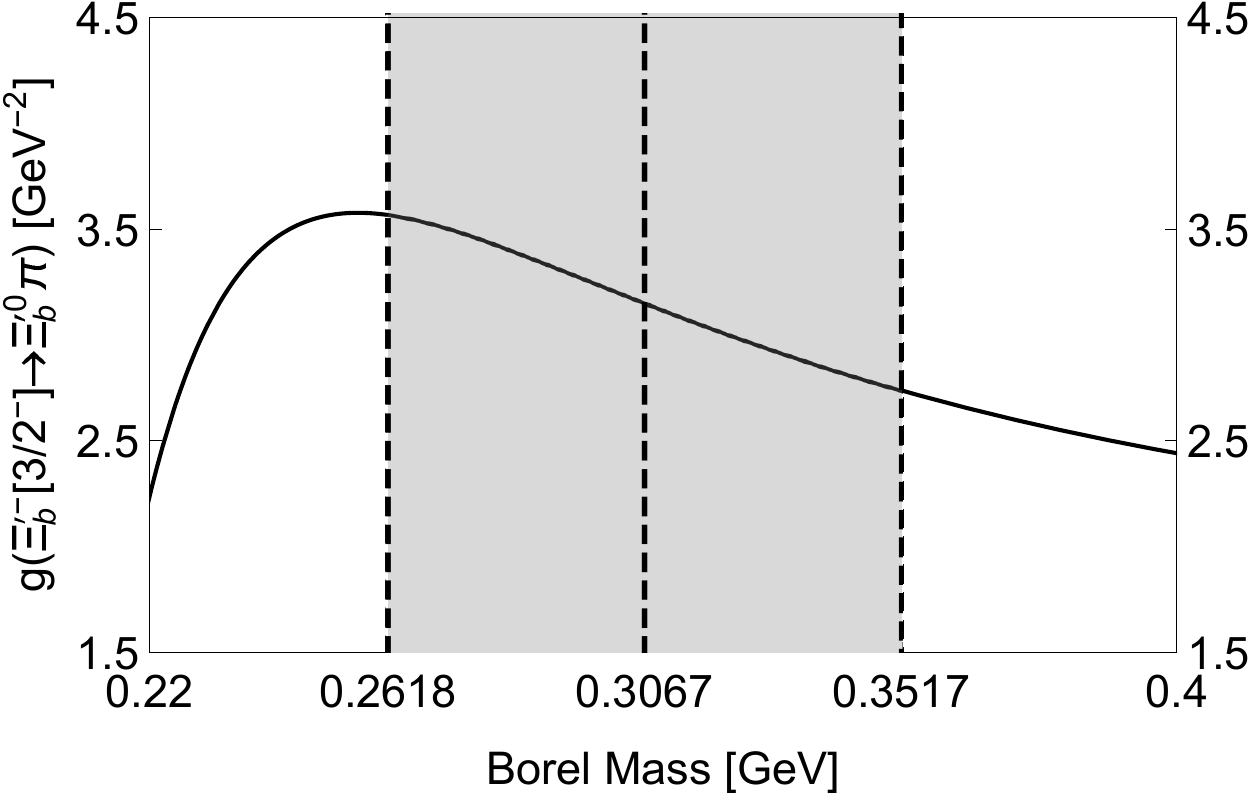}}}
\\
\subfigure[]{
\scalebox{0.5}{\includegraphics{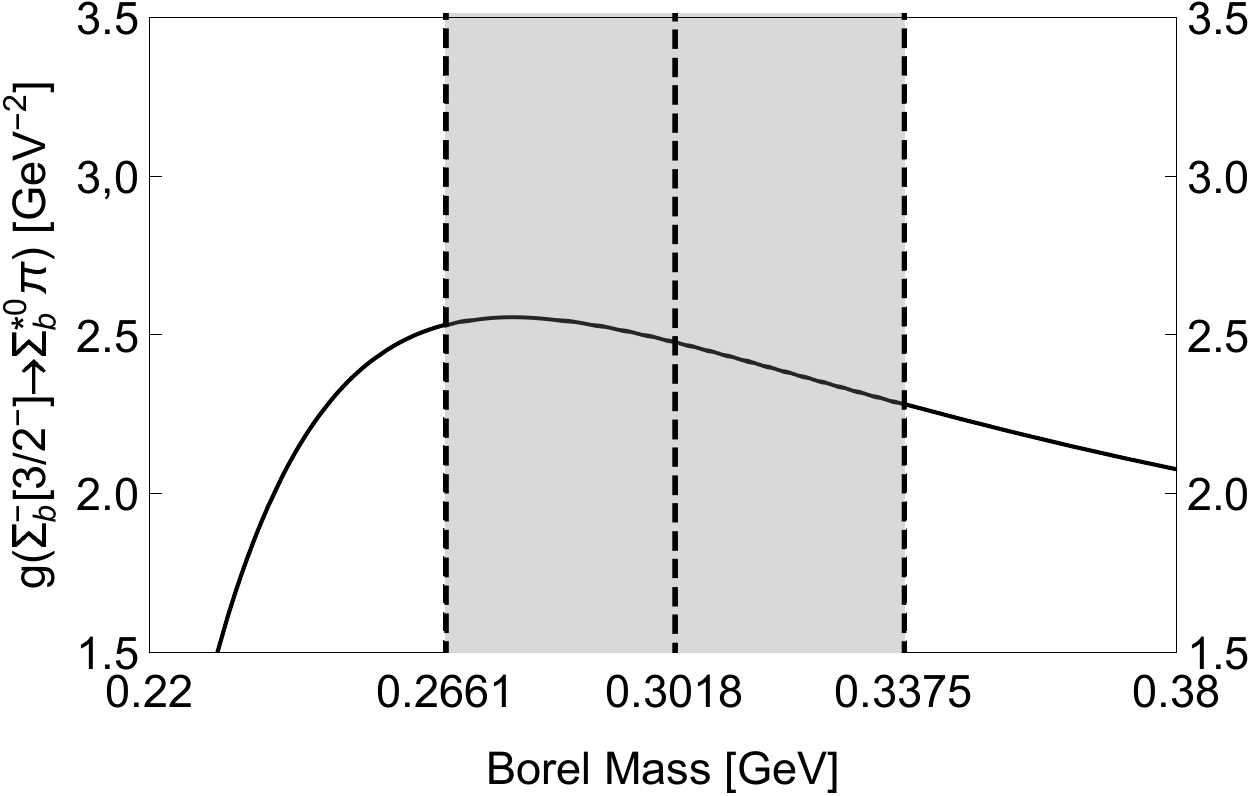}}}~~~~~
\subfigure[]{
\scalebox{0.5}{\includegraphics{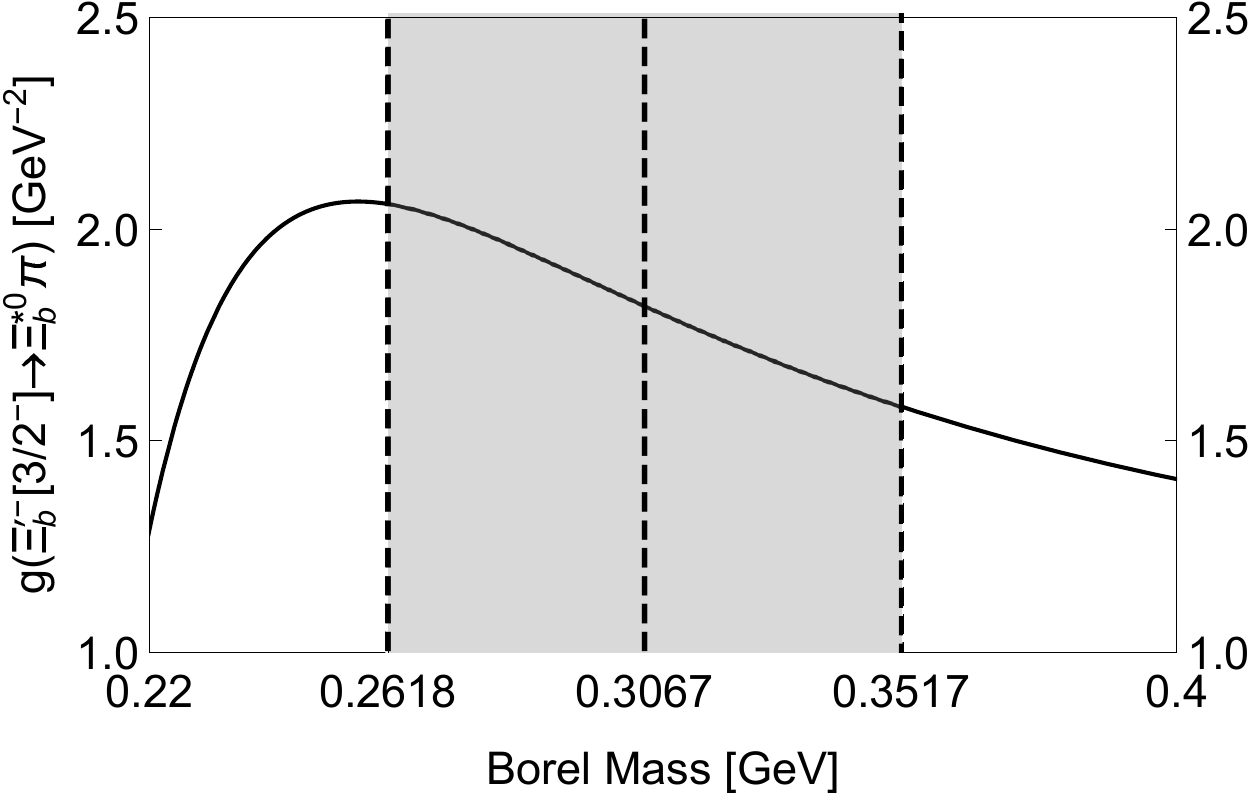}}}
\end{center}
\caption{The coupling constants as functions of the Borel mass $T$: (a) $g_{\Sigma_b^-[{3\over2}^-] \rightarrow \Lambda_b^0\pi^-}$, (b) $g_{\Xi_b^{\prime-}[{3\over2}^-] \rightarrow \Xi_b^0 \pi^-}$, (c) $g_{\Xi_b^{\prime-}[{3\over2}^-] \rightarrow \Lambda_b^0 K^-}$, (d) $g_{\Omega_b^-[{3\over2}^-] \rightarrow \Xi_b^0 K^-}$, (e) $g_{\Sigma_b^-[{3\over2}^-] \rightarrow \Sigma_b^0 \pi^-}$, (f) $g_{\Xi_b^{\prime-}[{3\over2}^-] \rightarrow \Xi_b^{\prime0} \pi^-}$, (g) $g_{\Sigma_b^-[{3\over2}^-] \rightarrow \Sigma_b^{*0} \pi^-}$, and (h) $g_{\Xi_b^{\prime-}[{3\over2}^-] \rightarrow \Xi_b^{*0} \pi^-}$.
Here the bottom baryon doublet $[\mathbf{6}_F, 2, 1, \lambda]$ is investigated.
\label{fig:621lambda}}
\end{figure*}

There are six bottom baryons contained in the $[\mathbf{6}_F, 2, 1, \lambda]$ doublet, that are $\Sigma_b({3\over2}^-/{5\over2}^-)$, $\Xi^\prime_b({3\over2}^-/{5\over2}^-)$, and $\Omega_b({3\over2}^-/{5\over2}^-)$. We study their $D$-wave decays into ground-state bottom baryons and pseudoscalar mesons, and find 24 non-zero coupling constants:
\begin{eqnarray}
\nonumber &(b1)& g_{\Sigma_b[{3\over2}^-]\to \Lambda_b[{1\over2}^+] \pi} = 7.57~{^{+4.50}_{-3.18}}~{\rm GeV}^{-2} \, ,
\\
\nonumber &(b2)& g_{\Sigma_b[{3\over2}^-]\to \Sigma_b[{1\over2}^+] \pi} = 4.29~{^{+3.13}_{-1.90}}~{\rm GeV}^{-2} \, ,
\\
\nonumber &(b3)& g_{\Sigma_b[{3\over2}^-]\to \Sigma_b^{*}[{3\over2}^+] \pi}= 2.48~{^{+1.48}_{-1.09}}~{\rm GeV}^{-2} \, ,
\\
\nonumber &(c1)& g_{\Sigma_b[{5\over2}^-]\to \Lambda_b[{1\over2}^+] \pi} = 7.57~{^{+4.50}_{-3.18}}~{\rm GeV}^{-2} \, ,
\\
\nonumber &(c2)& g_{\Sigma_b[{5\over2}^-]\to \Sigma_b[{1\over2}^+] \pi} = 2.86~{^{+2.09}_{-1.26}}~{\rm GeV}^{-2} \, ,
\\
\nonumber &(c3)& g_{\Sigma_b[{5\over2}^-]\to \Sigma_b^{*}[{3\over2}^+] \pi}= 2.20~{^{+1.31}_{-0.97}}~{\rm GeV}^{-2} \, ,
\\
\nonumber &(e1)& g_{\Xi_b^{\prime}[{3\over2}^-]\to \Xi_b[{1\over2}^+] \pi} = 4.92~{^{+2.68}_{-2.24}}~{\rm GeV}^{-2} \, ,
\\
\nonumber &(e2)& g_{\Xi_b^{\prime}[{3\over2}^-]\to \Lambda_b[{1\over2}^+] K} = 5.79~{^{+3.36}_{-2.38}}~{\rm GeV}^{-2} \, ,
\\
\nonumber &(e3)& g_{\Xi_b^{\prime}[{3\over2}^-]\to \Xi_b^{\prime}[{1\over2}^+] \pi } = 3.15~{^{+1.76}_{-1.37}}~{\rm GeV}^{-2} \, ,
\\
\nonumber &(e4)& g_{\Xi_b^{\prime}[{3\over2}^-]\to \Sigma_b[{1\over2}^+]K}= 2.43~{\rm GeV}^{-2} \, ,
\\
\nonumber &(e5)& g_{\Xi_b^{\prime}[{3\over2}^-]\to \Xi_b^{*}[{3\over2}^+]\pi}= 1.82~{^{+0.99}_{-0.76}}~{\rm GeV}^{-2} \, ,
\\
          &(e6)& g_{\Xi_b^{\prime}[{3\over2}^-]\to\Sigma_b^{*}[{3\over2}^+]K}= 1.40~{\rm GeV}^{-2} \, ,
\\
\nonumber &(f1)& g_{\Xi_b^{\prime}[{5\over2}^-]\to \Xi_b[{1\over2}^+] \pi} = 4.92~{^{+2.68}_{-2.24}}~{\rm GeV}^{-2} \, ,
\\
\nonumber &(f2)& g_{\Xi_b^{\prime}[{5\over2}^-]\to \Lambda_b[{1\over2}^+] K} = 5.79~{^{+3.36}_{-2.38}}~{\rm GeV}^{-2} \, ,
\\
\nonumber &(f3)& g_{\Xi_b^{\prime}[{5\over2}^-]\to \Xi_b^{\prime}[{1\over2}^+] \pi } = 2.10~{^{+1.18}_{-0.91}}~{\rm GeV}^{-2} \, ,
\\
\nonumber &(f4)& g_{\Xi_b^{\prime}[{5\over2}^-]\to \Sigma_b[{1\over2}^+]K}= 1.62~{\rm GeV}^{-2} \, ,
\\
\nonumber &(f5)& g_{\Xi_b^{\prime}[{5\over2}^-]\to \Xi_b^{*}[{3\over2}^+]\pi}= 1.62~{^{+0.88}_{-0.67}}~{\rm GeV}^{-2} \, ,
\\
\nonumber &(f6)& g_{\Xi_b^{\prime}[{5\over2}^-]\to\Sigma_b^{*}[{3\over2}^+]K}= 1.25~{\rm GeV}^{-2} \, ,
\\
\nonumber &(h1)& g_{\Omega_b[{3\over2}^-]\to \Xi_b[{1\over2}^+]K}= 7.27~{^{+3.92}_{-2.83}}~{\rm GeV}^{-2} \, ,
\\
\nonumber &(h2)& g_{\Omega_b[{3\over2}^-]\to \Xi_b^{\prime}[{1\over2}^+]K}= 3.78~{\rm GeV}^{-2} \, ,
\\
\nonumber &(h3)& g_{\Omega_b[{3\over2}^-]\to \Xi_b^{*}[{3\over2}^+]K}= 2.18~{\rm GeV}^{-2} \, ,
\\
\nonumber &(i1)& g_{\Omega_b[{5\over2}^-]\to \Xi_b[{1\over2}^+]K}= 7.27~{^{+3.92}_{-2.83}}~{\rm GeV}^{-2} \, ,
\\
\nonumber &(i2)& g_{\Omega_b[{5\over2}^-]\to \Xi_b^{\prime}[{1\over2}^+]K}= 2.52~{\rm GeV}^{-2} \, ,
\\
\nonumber &(i3)& g_{\Omega_b[{5\over2}^-]\to \Xi_b^{*}[{3\over2}^+]K}= 1.94~{\rm GeV}^{-2} \, .
\end{eqnarray}
We show some of these coupling constants as functions of the Borel mass $T$ in Fig.~\ref{fig:621lambda}. Based on them, we further find sixteen $D$-wave decay channels that are kinematically allowed:
\begin{eqnarray}
\nonumber &(b1)& \Gamma_{\Sigma_b[{3\over2}^-]\to \Lambda_b[{1\over2}^+] \pi} =49.6~{^{+76.4}_{-32.9}}~{\rm MeV} \, ,
\\
\nonumber &(b2)& \Gamma_{\Sigma_b[{3\over2}^-]\to \Sigma_b[{1\over2}^+] \pi} = 1.6~{^{+3.2}_{-1.1}}~{\rm MeV} \, ,
\\
\nonumber &(b3)& \Gamma_{\Sigma_b[{3\over2}^-]\to \Sigma_b^{*}[{3\over2}^+] \pi} = 0.23~{^{+0.36}_{-0.16}}~{\rm MeV} \, ,
\\
\nonumber &(c1)& \Gamma_{\Sigma_b[{5\over2}^-]\to \Lambda_b[{1\over2}^+] \pi} =20.8~{^{+23.7}_{-13.8}}~{\rm MeV} \, ,
\\
\nonumber &(c2)& \Gamma_{\Sigma_b[{5\over2}^-]\to \Sigma_b[{1\over2}^+] \pi} = 0.36~{^{+0.71}_{-0.24}}~{\rm MeV} \, ,
\\
\nonumber &(c3)& \Gamma_{\Sigma_b[{5\over2}^-]\to \Sigma_b^{*}[{3\over2}^+] \pi} = 1.1~{^{+1.8}_{-0.8}}~{\rm MeV} \, ,
\\
\nonumber &(e1)& \Gamma_{\Xi_b^{\prime}[{3\over2}^-]\to \Xi_b[{1\over2}^+] \pi}= 19.0~{^{+26.3}_{-13.3}}~{\rm MeV} \, ,
\\
          &(e2)& \Gamma_{\Xi_b^{\prime}[{3\over2}^-]\to \Lambda_b[{1\over2}^+] K}= 7.4~{^{+11.0}_{-~4.8}}~{\rm MeV} \, ,
\\
\nonumber &(e3)& \Gamma_{\Xi_b^{\prime}[{3\over2}^-]\to \Xi_b^{\prime}[{1\over2}^+] \pi} = 0.79~{^{+1.10}_{-0.79}}~{\rm MeV} \, ,
\\
\nonumber &(e5)& \Gamma_{\Xi_b^{\prime}[{3\over2}^-]\to \Xi_b^{*}[{3\over2}^+]\pi}= 0.12~{^{+0.17}_{-0.08}}~{\rm MeV } \, ,
\\
\nonumber &(f1)& \Gamma_{\Xi_b^{\prime}[{5\over2}^-]\to \Xi_b[{1\over2}^+] \pi}= 8.1~{^{+11.2}_{-~5.7}}~{\rm MeV} \, ,
\\
\nonumber &(f2)& \Gamma_{\Xi_b^{\prime}[{5\over2}^-]\to \Lambda_b[{1\over2}^+] K}= 3.4~{^{+5.1}_{-2.2}}~{\rm MeV} \, ,
\\
\nonumber &(f3)& \Gamma_{\Xi_b^{\prime}[{5\over2}^-]\to \Xi_b^{\prime}[{1\over2}^+] \pi} = 0.17~{^{+0.24}_{-0.11}}~{\rm MeV} \, ,
\\
\nonumber &(f5)& \Gamma_{\Xi_b^{\prime}[{5\over2}^-]\to \Xi_b^{*}[{3\over2}^+]\pi}= 0.58~{^{+0.80}_{-0.38}}~{\rm MeV } \, ,
\\
\nonumber &(h1)& \Gamma_{\Omega_b[{3\over2}^-]\to \Xi_b[{1\over2}^+]K}= 4.6~{^{+3.3}_{-1.9}}~{\rm MeV} \, ,
\\
\nonumber &(i1)& \Gamma_{\Omega_b[{5\over2}^-]\to \Xi_b[{1\over2}^+]K}= 2.5~{^{+3.5}_{-1.6}}~{\rm MeV} \, .
\end{eqnarray}
We summarize these $D$-wave decay widths in Table~\ref{tab:decay621lambda}, together with their mass spectrum, $S$-wave decay properties, and possible experimental candidates.

%
\section{Summary and Discussions}\label{sec:summary}
%

\begin{table*}[hbt]
\begin{center}
\renewcommand{\arraystretch}{1.5}
\caption{Decay properties of the $P$-wave bottom baryons belonging to the $[\mathbf{6}_F, 1, 0, \rho]$ doublet. In Refs.~\cite{Chen:2015kpa,Mao:2015gya} we studied the mass spectrum of $P$-wave bottom baryons, and the results are reanalysed and summarized in the second and third columns. In Refs.~\cite{Chen:2017sci,Yang:2019cvw} we studied the $S$-wave decay properties of $P$-wave bottom baryons into ground-state bottom baryons together with pseudoscalar mesons or vector mesons, and the results are reanalysed and summarized in the fifth column. The sixth column ``$D$-wave width'' is added in the present study. Possible experimental candidates are given in the last column for comparisons.}
\begin{tabular}{ c | c | c | c | c | c | c | c}
\hline\hline
Baryon & ~~~~Mass~~~~ & Difference & \multirow{2}{*}{~~~~~~~~~Decay channels~~~~~~~~~}  & ~$S$-wave width~  & ~$D$-wave width~ & ~Total width~ & \multirow{2}{*}{~Candidate~}
\\ ($j^P$) & ({GeV})& ({MeV}) & & ({MeV}) & ({MeV}) & ({MeV})
\\ \hline\hline
\multirow{4}{*}{$\Sigma_b({1\over2}^-)$}&\multirow{4}{*}{$6.05 \pm 0.12$}&\multirow{6}{*}{$3\pm1$}&$\Sigma_b({1\over2}^-)\to \Sigma_b\pi$&$710$&--&\multirow{3}{*}{$710$}&\multirow{3}{*}{--}
\\ \cline{4-6}
&&&$\Sigma_b({1\over2}^-)\to \Sigma_b^{*} \pi$  & --  & $ 0.62$ &
\\ \cline{4-6}
&&&$\Sigma_b({1\over2}^-)\to \Lambda_b\rho\to \Lambda_b\pi\pi$& \multicolumn{2}{c|}{$4.3\times 10^{-3}$ }&
\\ \cline{1-2} \cline{4-8}
\multirow{4}{*}{$\Sigma_b({3\over2}^-)$}&\multirow{4}{*}{$6.05 \pm 0.12$}&&$\Sigma_b({3\over2}^-)\to \Sigma_b \pi$ & -- & $0.84$&\multirow{3}{*}{$410$}&\multirow{3}{*}{--}
\\ \cline{4-6}
&&&$\Sigma_b({3\over2}^-)\to \Sigma_b^{*} \pi$ & $410$& $ 0.098$&
\\ \cline{4-6}
&&&$\Sigma_b({3\over2}^-)\to \Lambda_b\rho\to \Lambda_b\pi\pi$&\multicolumn{2}{c|}{$5.1\times 10^{-3}$}&
\\ \hline
\multirow{3}{*}{$\Xi^\prime_b({1\over2}^-)$}&\multirow{3}{*}{$6.18 \pm 0.12$}&\multirow{6}{*}{$3 \pm 1$}&$\Xi_b^{\prime}({1\over2})\to \Xi_b^{\prime}\pi$& $250$&--&\multirow{3}{*}{250}&\multirow{3}{*}{--}
\\ \cline{4-6}
&&&$\Xi_b^{\prime}({1\over2}^-)\to\Xi_b^{*}\pi$&--&$0.29$&
\\ \cline{4-6}
&&&$\Xi_b^{\prime}({1\over2}^-)\to\Xi_b\rho\to\Xi_b\pi\pi$&\multicolumn{2}{c|}{$1.2\times10^{-5}$}&
\\ \cline{1-2} \cline{4-8}
\multirow{3}{*}{$\Xi_b^{\prime}({3\over2}^-)$}&\multirow{3}{*}{$6.19 \pm 0.11$}&&$\Xi_b^{\prime}({3\over2}^-)\to\Xi_ b^{\prime}\pi$&--&$0.47$&\multirow{3}{*}{160}&\multirow{3}{*}{--}
\\ \cline{4-6}
&&&$\Xi_b^{\prime}({3\over2}^-)\to\Xi_b^{*}\pi$&$160$&$0.064$&
\\ \cline{4-6}
&&&$\Xi_b^{\prime}({3\over2}^-)\to \Xi_b\rho\to\Xi_b\pi\pi$&\multicolumn{2}{c|}{$8.0\times10^{-5}$}&
\\ \hline
$\Omega_b({1\over2}^-)$&$6.32 \pm 0.11$&\multirow{2}{*}{$2 \pm 1$}&\multicolumn{3}{c|}{--}&$\sim~0$&\multirow{2}{*}{$\Omega_b(6316)^-$~\cite{Aaij:2020cex}}
\\ \cline{1-2} \cline{4-7}
$\Omega_b({3\over2}^-)$&$6.32 \pm 0.11$&&\multicolumn{3}{c|}{--}&$\sim~0$
\\ \hline\hline
\end{tabular}
\label{tab:decay610rho}
\end{center}
\end{table*}

\begin{table*}[hbt]
\begin{center}
\renewcommand{\arraystretch}{1.5}
\caption{Decay properties of the $P$-wave bottom baryons belonging to the $[\mathbf{6}_F,0,1,\lambda]$ singlet. See the caption of Table~\ref{tab:decay610rho} for detailed explanations.}
\begin{tabular}{c|c|c|c|c|c|c|c}
\hline\hline
Baryon & ~~~~Mass~~~~ & Difference & \multirow{2}{*}{~~~~~~~~~Decay channels~~~~~~~~~}  & ~$S$-wave width~  & ~$D$-wave width~ & ~Total width~ & \multirow{2}{*}{~Candidate~}
\\ ($j^P$) & ({GeV})& ({MeV}) & & ({MeV}) & ({MeV}) & ({MeV})
\\ \hline\hline
$\Sigma_b({1\over2}^-)$&$6.05 \pm 0.11$&--&$\Sigma_b({1\over2}^-)\to\Lambda_b\pi$&$1300$&--&$1300$&--
\\ \hline
\multirow{2}{*}{$\Xi_b^{\prime}({1\over2}^-)$}&\multirow{2}{*}{$6.20 \pm 0.11$}&\multirow{2}{*}{--}&$\Xi_b^{\prime}({1\over2}^-)\to\Xi_b\pi$&$990$&--&\multirow{2}{*}{$1900$}&\multirow{2}{*}{--}
\\ \cline{4-6}
&&&$\Xi_b^{\prime}({1\over2}^-)\to\Lambda_b K$&$910$&--&
\\ \hline
$\Omega_b({1\over2}^-)$&$6.34\pm0.11$&--&$\Omega_b({1\over2}^-)\to\Xi_b K$&$2700$&--&$2700$&--
\\ \hline\hline
\end{tabular}
\label{tab:decay601lambda}
\end{center}
\end{table*}

\begin{table*}[hbt]
\begin{center}
\renewcommand{\arraystretch}{1.5}
\caption{Decay properties of the $P$-wave bottom baryons belonging to the $[\mathbf{6}_F, 1, 1, \lambda]$ doublet. See the caption of Table~\ref{tab:decay610rho} for detailed explanations.}
\begin{tabular}{ c | c | c | c | c | c | c | c}
\hline\hline
Baryon & ~~~~Mass~~~~ & Difference & \multirow{2}{*}{~~~~~~~~~Decay channels~~~~~~~~~}  & ~$S$-wave width~  & ~$D$-wave width~ & ~Total width~ & \multirow{2}{*}{~Candidate~}
\\ ($j^P$) & ({GeV})& ({MeV}) & & ({MeV}) & ({MeV}) & ({MeV})
\\ \hline\hline
\multirow{4}{*}{$\Sigma_b({1\over2}^-)$}&\multirow{4}{*}{$6.06 \pm 0.13$}&\multirow{6}{*}{$6\pm3$}&$\Sigma_b({1\over2}^-)\to \Sigma_b\pi$&$14.1~{^{+21.2}_{-10.9}}$&--&\multirow{3}{*}{$14.3~{^{+21.2}_{-10.9}}$}&\multirow{3}{*}{--}
\\ \cline{4-6}
&&&$\Sigma_b({1\over2}^-)\to \Sigma_b^{*} \pi$  & --  & $ 0.076~{^{+0.144}_{-0.076}}$ &
\\ \cline{4-6}
&&&$\Sigma_b({1\over2}^-)\to \Lambda_b\rho\to \Lambda_b\pi\pi$& \multicolumn{2}{c|}{$0.087~{^{+0.224}_{-0.085}}$ }&
\\ \cline{1-2} \cline{4-8}
\multirow{4}{*}{$\Sigma_b({3\over2}^-)$}&\multirow{4}{*}{$6.07 \pm 0.13$}&&$\Sigma_b({3\over2}^-)\to \Sigma_b \pi$ & -- & $0.55~{^{+0.74}_{-0.36}}$&\multirow{3}{*}{$4.8~{^{+5.9}_{-2.9}}$}&\multirow{3}{*}{--}
\\ \cline{4-6}
&&&$\Sigma_b({3\over2}^-)\to \Sigma_b^{*} \pi$ & $3.9~{^{+5.8}_{-2.9}}$& $ 0.070~{^{+0.096}_{-0.047}}$&
\\ \cline{4-6}
&&&$\Sigma_b({3\over2}^-)\to \Lambda_b\rho\to \Lambda_b\pi\pi$&\multicolumn{2}{c|}{$0.23~{^{+0.45}_{-0.20}}$}&
\\ \hline
\multirow{3}{*}{$\Xi^\prime_b({1\over2}^-)$}&\multirow{3}{*}{$6.21 \pm 0.11$}&\multirow{6}{*}{$7 \pm 2$}&$\Xi_b^{\prime}({1\over2})\to \Xi_b^{\prime}\pi$& $4.5~{^{+5.8}_{-3.3}}$&--&\multirow{3}{*}{$4.7~{^{+5.8}_{-3.3}}$}&\multirow{3}{*}{--}
\\ \cline{4-6}
&&&$\Xi_b^{\prime}({1\over2}^-)\to\Xi_b^{*}\pi$&--&$0.16~{^{+0.18}_{-0.10}}$&
\\ \cline{4-6}
&&&$\Xi_b^{\prime}({1\over2}^-)\to\Xi_b\rho\to\Xi_b\pi\pi$&\multicolumn{2}{c|}{$0.043~{^{+0.079}_{-0.038}}$}&
\\ \cline{1-2} \cline{4-8}
\multirow{4}{*}{$\Xi_b^{\prime}({3\over2}^-)$}&\multirow{4}{*}{$6.22 \pm 0.11$}&&$\Xi_b^{\prime}({3\over2}^-)\to\Xi_ b^{\prime}\pi$&--&$0.34~{^{+0.35}_{-0.20}}$&\multirow{4}{*}{$1.8~{^{+1.1}_{-1.0}}$}&\multirow{4}{*}{--}
\\ \cline{4-6}
&&&$\Xi_b^{\prime}({3\over2}^-)\to\Xi_b^{*}\pi$&$1.3~{^{+1.0}_{-0.9}}$&$0.051~{^{+0.057}_{-0.030}}$&
\\ \cline{4-6}
&&&$\Xi_b^{\prime}({3\over2}^-)\to \Xi_b\rho\to\Xi_b\pi\pi$&\multicolumn{2}{c|}{$0.078~{^{+0.147}_{-0.068}}$}&
\\ \cline{4-6}
&&&$\Xi_b^{\prime}({3\over2}^-)\to\Xi_b^{\prime}\rho\to\Xi_b^{\prime}\pi\pi$&\multicolumn{2}{c|}{$(5.5~{^{+6.4}_{-3.5}})\times10^{-6}$}&
\\ \hline
$\Omega_b({1\over2}^-)$&$6.34 \pm 0.10$&\multirow{2}{*}{$6 \pm 2$}&\multicolumn{3}{c|}{--}&$\sim~0$&$\Omega_b(6330)^-$~\cite{Aaij:2020cex}
\\ \cline{1-2} \cline{4-8}
$\Omega_b({3\over2}^-)$&$6.34 \pm 0.09$&&\multicolumn{3}{c|}{--}&$\sim~0$&$\Omega_b(6340)^-$~\cite{Aaij:2020cex}
\\ \hline\hline
\end{tabular}
\label{tab:decay611lambda}
\end{center}
\end{table*}

\begin{table*}[hbt]
\begin{center}
\renewcommand{\arraystretch}{1.5}
\caption{Decay properties of the $P$-wave bottom baryons belonging to the $[\mathbf{6}_F, 2, 1, \lambda]$ doublet. See the caption of Table~\ref{tab:decay610rho} for detailed explanations.}
\begin{tabular}{ c | c | c | c | c | c | c | c}
\hline\hline
Baryon & ~~~~Mass~~~~ & Difference & \multirow{2}{*}{~~~~~~~Decay channels~~~~~~~}  & ~$S$-wave width~  & ~$D$-wave width~ & ~Total width~ & \multirow{2}{*}{~Candidate~}
\\ ($j^P$) & ({GeV})& ({MeV}) & & ({MeV}) & ({MeV}) & ({MeV})
\\ \hline\hline
\multirow{4}{*}{$\Sigma_b({3\over2}^-)$}&\multirow{4}{*}{$6.11\pm0.16$}&\multirow{5}{*}{$12\pm5$}&$\Sigma_b({3\over2}^-)\to\Lambda_b\pi$&--&$49.6~{^{+76.4}_{-32.9}}$&\multirow{4}{*}{$51.4~{^{+76.5}_{-32.9}}$}&\multirow{4}{*}{$\Sigma_b(6097)^\pm$~\cite{Aaij:2018tnn}}
\\ \cline{4-6}
&&&$\Sigma_b({3\over2}^-)\to \Sigma_b \pi$ &--& $1.6~{^{+3.2}_{-1.1}}$&
\\ \cline{4-6}
&&&$\Sigma_b({3\over2}^-)\to \Sigma_b^{*}\pi$ & $0.019~{^{+0.065}_{-0.019}}$ & $ 0.23~{^{+0.36}_{-0.16}}$&
\\ \cline{4-6}
&&&$\Sigma_b({3\over2}^-)\to\Sigma_b\rho\to\Sigma_b\pi\pi$&\multicolumn{2}{c|}{$(1.4~{^{+2.5}_{-1.1}})\times10^{-4}$}&
\\ \cline{1-2} \cline{4-8}
\multirow{3}{*}{$\Sigma_b({5\over2}^-)$}&\multirow{3}{*}{$6.12\pm0.15$}&&$\Sigma_b({5\over2}^-)\to \Lambda_b \pi$ &--& $20.8~{^{+23.7}_{-13.8}}$&\multirow{3}{*}{$22.3~{^{+23.8}_{-13.8}}$}&\multirow{3}{*}{--}
\\ \cline{4-6}
&&&$\Sigma_b({5\over2}^-)\to \Sigma_b \pi$ &--& $0.36~{^{+0.71}_{-0.24}}$&
\\ \cline{4-6}
&&&$\Sigma_b({5\over2}^-)\to \Sigma_b^{*}\pi$ &--& $1.1~{^{+1.8}_{-0.8}}$ &
\\ \hline
\multirow{5}{*}{$\Xi_b^{\prime}({3\over2}^-)$}&\multirow{5}{*}{$6.23\pm0.15$}&\multirow{6}{*}{$11\pm5$}&$\Xi_b^{\prime}({3\over2}^-)\to \Xi_b\pi$ &--& $19.0~{^{+26.3}_{-13.3}}$ & \multirow{5}{*}{$27.3~{^{+28.5}_{-14.2}}$}&\multirow{5}{*}{$\Xi_b(6227)^-$~\cite{Aaij:2018yqz}}
\\ \cline{4-6}
&&&$\Xi_b^{\prime}({3\over2}^-)\to \Lambda_bK$ &--& $7.4~{^{+11.0}_{-~4.8}}$&
\\ \cline{4-6}
&&&$\Xi_b^{\prime}({3\over2}^-)\to \Xi_b^{\prime}\pi$ &--&$0.79~{^{+1.13}_{-0.79}}$&
\\ \cline{4-6}
&&&$\Xi_b^{\prime}({3\over2}^-)\to \Xi_b^{*}\pi$ &$0.007~{^{+0.023}_{-0.007}}$&$0.12~{^{+0.17}_{-0.08}}$&
\\ \cline{4-6}
&&&$\Xi_b^{\prime}({3\over2}^-)\to\Xi_b^{\prime}\rho\to\Xi_b^{\prime}\pi\pi$&\multicolumn{2}{c|}{$ (5.6~{^{+9.1}_{-4.3}})\times10^{-4}$}&
\\ \cline{1-2} \cline{4-8}
\multirow{5}{*}{$\Xi^\prime_b({5\over2}^-)$}&\multirow{5}{*}{$6.24\pm0.14$}&&$\Xi_b^{\prime}({5\over2}^-)\to \Lambda_bK$ &--& $3.4~{^{+5.1}_{-2.2}}$&\multirow{5}{*}{$12.3~{^{+12.3}_{-~6.1}}$}&\multirow{5}{*}{--}
\\ \cline{4-6}
&&&$\Xi_b^{\prime}({5\over2}^-)\to \Xi_b \pi$&--&$8.1~{^{+11.2}_{-~5.7}}$&
\\ \cline{4-6}
&&&$\Xi_b^{\prime}({5\over2}^-)\to \Xi_b^{\prime}\pi$ &--&$0.17~{^{+0.24}_{-0.11}}$&
\\ \cline{4-6}
&&&$\Xi_b^{\prime}({5\over2}^-)\to \Xi_b^{*}\pi$ &--&$0.58~{^{+0.80}_{-0.38}}$&
\\ \cline{4-6}
&&&$\Xi_b^{\prime}({5\over2}^-)\to\Xi_b^{*}\rho\to\Xi_b^{*}\pi\pi$&\multicolumn{2}{c|}{$(6.4~{^{+10.3}_{-~4.5}})\times10^{-5}$}&
\\ \hline
$\Omega_b({3\over2}^-)$&$6.35\pm0.13$&\multirow{2}{*}{$10\pm4$}&$\Omega_b({3\over2}^-)\to \Xi_b K$ &--& $4.6~{^{+3.3}_{-1.9}}$&$4.6~{^{+3.3}_{-1.9}}$&$\Omega_b(6350)^-$~\cite{Aaij:2020cex}
\\ \cline{1-2} \cline{4-8}
$\Omega_b({5\over2}^-)$&$6.36\pm0.12$&&$\Omega_b({5\over2}^-)\to \Xi_b K$ &--&$2.5~{^{+3.5}_{-1.6}}$& $2.5~{^{+3.5}_{-1.6}}$ &--
\\ \hline\hline
\end{tabular}
\label{tab:decay621lambda}
\end{center}
\end{table*}

To summarize this paper, we have investigated the $P$-wave bottom baryons belonging to the $SU(3)$ flavor $\mathbf{6}_F$ representation, and studied their $D$-wave decays into ground-state bottom baryons and pseudoscalar mesons. Together with Refs.~\cite{Chen:2015kpa,Mao:2015gya,Chen:2017sci,Yang:2019cvw}, we have performed a rather complete study on both mass spectra and decay properties of $P$-wave bottom baryons using the method of QCD sum rules and light-cone sum rules within the framework of heavy quark effective theory.

Accordingly to the heavy quark effective theory, we categorize the $P$-wave bottom baryons of the $SU(3)$ flavor $\mathbf{6}_F$ into four multiplets: $[\mathbf{6}_F, 1, 0, \rho]$, $[\mathbf{6}_F, 0, 1, \lambda]$, $[\mathbf{6}_F, 1, 1, \lambda]$, and $[\mathbf{6}_F, 2, 1, \lambda]$. In this paper we have studied their $D$-wave decay properties, and the results are separately summarized in Tables~\ref{tab:decay610rho}/\ref{tab:decay601lambda}/\ref{tab:decay611lambda}/\ref{tab:decay621lambda}. Besides, in Refs.~\cite{Chen:2015kpa,Mao:2015gya} we have studied the mass spectrum of $P$-wave bottom baryons, and the results are reanalysed and summarized in these tables; in Refs.~\cite{Chen:2017sci,Yang:2019cvw} we have studied $S$-wave decay properties of $P$-wave bottom baryons into ground-state bottom baryons together with pseudoscalar mesons or vector mesons, and the results are also reanalysed and summarized in these tables.

\begin{figure*}[htb]
\begin{center}
\scalebox{0.75}{\includegraphics{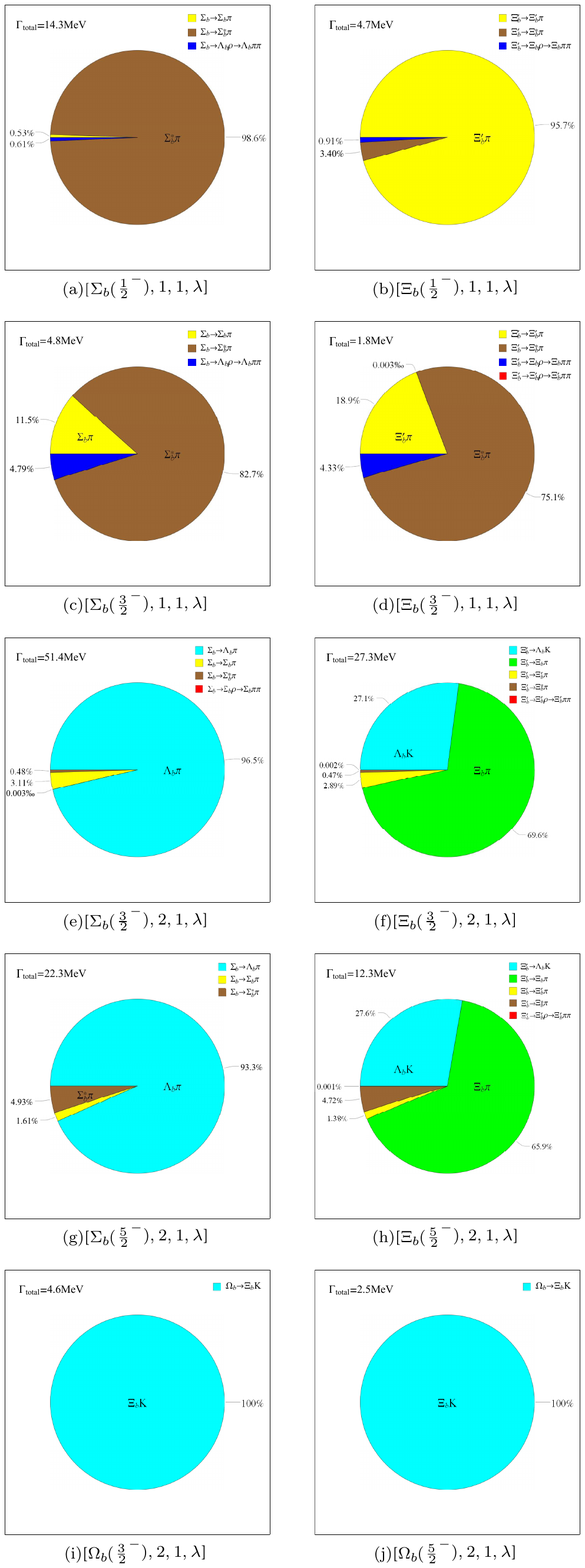}}
\end{center}
\caption{Branching ratios of eight $P$-wave bottom baryons of the flavor $\mathbf{6}_F$, with limited but non-zero widths ($0< \Gamma < 100$~MeV). See the caption of Table~\ref{tab:result} for detailed explanations.
\label{fig:pie}}
\end{figure*}

Before drawing our conclusions, we note that there are considerable (theoretical) uncertainties in our results for the absolute values of the bottom baryon masses due to their significant dependence on the bottom quark mass~\cite{Chen:2015kpa,Mao:2015gya}; however, their mass splittings within the same doublets do not depend much on this, so they are produced quite well with much less (theoretical) uncertainties and give more useful information; moreover, we can extract even (much) more useful information from $S$- and $D$-wave strong decay properties of $P$-wave bottom baryons. Based on the results summarized in Tables~\ref{tab:decay610rho}/\ref{tab:decay601lambda}/\ref{tab:decay611lambda}/\ref{tab:decay621lambda}, we can well understand $P$-wave bottom baryons as a whole:
\begin{itemize}

\item The $[\mathbf{6}_F,0,1,\lambda]$ singlet contains three bottom baryons: $\Sigma_b({1\over2}^-)$, $\Xi^\prime_b({1\over2}^-)$, and $\Omega_b({1\over2}^-)$. Their total widths are all calculated to be very large, preventing them to be observed in any experiment.

\item The $[\mathbf{6}_F, 1, 0, \rho]$ doublet contains six bottom baryons: $\Sigma_b({1\over2}^-/{3\over2}^-)$, $\Xi^\prime_b({1\over2}^-/{3\over2}^-)$, and $\Omega_b({1\over2}^-/{3\over2}^-)$. The total widths of $\Sigma_b({1\over2}^-/{3\over2}^-)$ and $\Xi^\prime_b({1\over2}^-/{3\over2}^-)$ are all calculated to be very large, while the total widths of $\Omega_b({1\over2}^-/{3\over2}^-)$ are both extracted to be zero.

\item The $[\mathbf{6}_F, 1, 1, \lambda]$ doublet contains six bottom baryons: $\Sigma_b({1\over2}^-/{3\over2}^-)$, $\Xi^\prime_b({1\over2}^-/{3\over2}^-)$, and $\Omega_b({1\over2}^-/{3\over2}^-)$. Their total widths are all calculated to be less than 100~MeV.

\item The $[\mathbf{6}_F, 2, 1, \lambda]$ doublet contains six bottom baryons: $\Sigma_b({3\over2}^-/{5\over2}^-)$, $\Xi^\prime_b({3\over2}^-/{5\over2}^-)$, and $\Omega_b({3\over2}^-/{5\over2}^-)$. Their total widths are all calculated to be less than 100~MeV.

\end{itemize}
Hence, among all the possible $P$-wave bottom baryons of the flavor $\mathbf{6}_F$, we find altogether four $\Sigma_b$, four $\Xi^\prime_b$, and six $\Omega_b$ baryons, with limited widths ($< 100$~MeV) and so capable of being observed. Their masses, mass splittings within the same multiplets, and decay properties are summarized in Table~\ref{tab:result}. Their possible experimental candidates are also given in this table for comparisons. Among these fourteen bottom baryons, ten of them have non-zero decay widths, whose branching ratios are shown in Fig.~\ref{fig:pie} using pie charts. We suggest the LHCb and CMS Collaborations to search for these excited bottom baryons, but note that it still depends on the production rates whether these baryons can be observed or not. Especially, it is interesting to further investigate the $\Lambda_b(6072)^0$, {\it i.e.}, the broad excess of events in the $\Lambda_b^0 \pi^+ \pi^-$ mass distribution in the region of $6040$-$6100$~MeV~\cite{Sirunyan:2020gtz,Aaij:2020rkw}.

In the present study the $\rho$-mode doublet $[\mathbf{6}_F, 1, 0, \rho]$ is found to be lower than the two $\lambda$-mode doublets $[{\bf 6}_F, 1, 1, \lambda]$ and $[{\bf 6}_F, 2, 1, \lambda]$, a behaviour which is consistent with our previous results for their corresponding doublets of the $SU(3)$ flavor $\mathbf{\bar 3}_F$~\cite{Chen:2015kpa,Mao:2015gya}, but in contrast to the quark model expectation~\cite{Copley:1979wj,Yoshida:2015tia}. However, this is possible simply because
the mass differences between different multiplets have considerable uncertainties in our framework, similar to the absolute values of baryon masses, but unlike the mass differences within the same multiplet. We propose to verify whether the $\rho$-mode doublet $[\mathbf{6}_F, 1, 0, \rho]$ exist or not by investigating: a) the spin-parity quantum number of the $\Omega_b(6316)^-$, b) whether it can be separated into two states almost degenerate, and c) whether its $\Sigma_b$ and $\Xi^\prime_b$ partner states can be observed.

\begin{table*}[hbt]
\begin{center}
\renewcommand{\arraystretch}{1.4}
\caption{Among all the possible $P$-wave bottom baryons of the flavor $\mathbf{6}_F$, we find altogether four $\Sigma_b$, four $\Xi^\prime_b$, and six $\Omega_b$ baryons, with limited widths ($< 100$~MeV) and so capable of being observed. Their masses, mass splittings within the same multiplets, and decay properties are extracted for future experimental searches. We note that there are considerable uncertainties in our results for the absolute values of the bottom baryon masses due to their significant dependence on the bottom quark mass~\cite{Chen:2015kpa,Mao:2015gya}; however, their mass splittings within the same doublets do not depend much on this, so they are produced quite well with much less uncertainties and give more useful information; moreover, we can extract even more useful information from $S$- and $D$-wave strong decay properties of $P$-wave bottom baryons.}
\begin{tabular}{ c | c | c | c | c | c | c | c}
\hline\hline
\multirow{2}{*}{~B~} & \multirow{2}{*}{~Multiplet~} & ~Baryon~ & ~~~Mass~~~ & Difference & \multirow{2}{*}{~~~~~~~~~~~Decay channel~~~~~~~~~~~ }& Total width  & \multirow{2}{*}{~Candidate~}
\\&  & ($j^P$) & ({GeV}) & ({MeV}) & & ({MeV}) &
\\ \hline\hline
\multirow{9}{*}{$\Sigma_b$}
& \multirow{4}{*}{$[\mathbf{6}_F, 1, 1, \lambda]$} & $\Sigma_b({1\over2}^-)$ & $6.06\pm0.13$& \multirow{4}{*}{$6\pm3$} &
$\begin{array}{c}
\Gamma\left(\Sigma_b({1\over2}^-)\to \Sigma_b\pi\right)=14.1~{^{+21.2}_{-10.9}}~{\rm MeV} \\
\Gamma\left(\Sigma_b({1\over2}^-)\to \Sigma_b^{*}\pi\right)=0.076~{^{+0.144}_{-0.076}}~{\rm MeV}\\
\Gamma\left(\Sigma_b({1\over2}^-)\to \Lambda_b\rho \to \Lambda_b\pi\pi\right)=0.087~{\rm MeV}
\end{array}$ &$14.3~{^{+21.2}_{-10.9}}$&--
\\ \cline{3-4}\cline{6-8}
                                                                           && $\Sigma_b({3\over2}^-)$ & $6.07\pm0.13$ &&
$\begin{array}{c}
\Gamma\left(\Sigma_b({3\over2}^-)\to \Sigma_b\pi\right)=0.55~{^{+0.74}_{-0.36}}~{\rm MeV} \\
\Gamma\left(\Sigma_b({3\over2}^-)\to \Sigma_b^{*}\pi\right)=4.0~{^{+5.8}_{-2.9}}~{\rm MeV} \\
\Gamma\left(\Sigma_b({3\over2}^-)\to \Lambda_b\rho \to \Lambda_b\pi\pi\right)=0.23~{\rm MeV}
\end{array}$  &$4.8~{^{+5.9}_{-2.9}}$&--
\\ \cline{2-8}
&\multirow{4}{*}{$[\mathbf{6}_F,2,1,\lambda]$}&$\Sigma_b({3\over2}^-)$&$6.11\pm0.16$&\multirow{2}{*}{$12\pm5$}&
$\begin{array}{c}
\Gamma\left(\Sigma_b({3\over2}^-)\to\Lambda_b\pi\right)=49.6~{^{+76.4}_{-32.9}}~{\rm MeV}\\
\Gamma\left(\Sigma_b({3\over2}^-)\to \Sigma_b \pi\right)=1.6~{^{+3.2}_{-1.1}}~{\rm MeV}\\
\Gamma\left(\Sigma_b({3\over2}^-)\to \Sigma_b^{*}\pi\right)=0.25~{^{+0.37}_{-0.16}}~{\rm MeV}\\
\Gamma\left(\Sigma_b({3\over2}^-)\to\Sigma_b\rho\to\Sigma_b\pi\pi\right)=1.4\times10^{-4}~{\rm MeV}
\end{array}$
&$51.4~{^{+76.5}_{-32.9}}$&$\Sigma_b(6097)^\pm$
\\ \cline{3-4} \cline{6-8}
&&$\Sigma_b({5\over2}^-)$&$6.12\pm0.15$&&$\begin{array}{c}
\Gamma\left(\Sigma_b({5\over2}^-)\to\Lambda_b\pi\right)=20.8~{^{+23.7}_{-13.8}}~{\rm MeV}\\
\Gamma\left(\Sigma_b({5\over2}^-)\to \Sigma_b \pi\right)=0.36~{^{+0.71}_{-0.24}}~{\rm MeV}\\
\Gamma\left(\Sigma_b({5\over2}^-)\to \Sigma_b^{*}\pi\right)=1.1~{^{+1.8}_{-0.8}}~{\rm MeV}
\end{array}$&$22.3~{^{+23.8}_{-13.8}}$&--
\\ \hline \hline
\multirow{10}{*}{$\Xi^\prime_b$}
&\multirow{5}{*}{$[\mathbf{6}_F,1,1,\lambda]$}&$\Xi^\prime_b({1\over2}^-)$&$6.21 \pm 0.11$&\multirow{5}{*}{$7 \pm 2$}&
$\begin{array}{c}
\Gamma\left(\Xi_b^{\prime}({1\over2})\to \Xi_b^{\prime}\pi\right)=4.5~{^{+5.8}_{-3.3}}~{\rm MeV}\\
\Gamma\left(\Xi_b^{\prime}({1\over2}^-)\to\Xi_b^{*}\pi\right)=0.16~{^{+0.18}_{-0.10}}~{\rm MeV}\\
\Gamma\left(\Xi_b^{\prime}({1\over2}^-)\to\Xi_b\rho\to\Xi_b\pi\pi\right)=0.043~{\rm MeV}
\end{array}$&$4.7~{^{+5.8}_{-3.3}}$&--
\\ \cline{3-4}\cline{6-8}
&&$\Xi_b^{\prime}({3\over2}^-)$&$6.22 \pm 0.11$&&
$\begin{array}{c}
\Gamma\left(\Xi_b^{\prime}({3\over2}^-)\to\Xi_ b^{\prime}\pi\right)=0.34~{^{+0.35}_{-0.20}}~{\rm MeV}\\
\Gamma\left(\Xi_b^{\prime}({3\over2}^-)\to\Xi_b^{*}\pi\right)=1.4~{^{+1.0}_{-0.9}}~{\rm MeV}\\
\Gamma\left(\Xi_b^{\prime}({3\over2}^-)\to \Xi_b\rho\to\Xi_b\pi\pi\right)=0.078~{\rm MeV}\\
\Gamma\left(\Xi_b^{\prime}({3\over2}^-)\to\Xi_b^{\prime}\rho\to\Xi_b^{\prime}\pi\pi\right)=5.5\times10^{-6}~{\rm MeV}
\end{array}$&$1.8~{^{+1.1}_{-1.0}}$&--
\\ \cline{2-8}
&\multirow{6}{*}{$[\mathbf{6}_F,2,1,\lambda]$}&$\Xi_b^{\prime}({3\over2}^-)$&$6.23\pm0.15$&\multirow{3}{*}{$11\pm5$}&
$\begin{array}{c}
\Gamma\left(\Xi_b^{\prime}({3\over2}^-)\to \Xi_b\pi\right)=19.0~{^{+26.3}_{-13.3}}~{\rm MeV}\\
\Gamma\left(\Xi_b^{\prime}({3\over2}^-)\to \Lambda_b K\right)=7.4~{^{+11.0}_{-~4.8}}~{\rm MeV}\\
\Gamma\left(\Xi_b^{\prime}({3\over2}^-)\to \Xi_b^{\prime}\pi\right)=0.79~{^{+1.13}_{-0.79}}~{\rm MeV}\\
\Gamma\left(\Xi_b^{\prime}({3\over2}^-)\to \Xi_b^{*}\pi\right)=0.13~{^{+0.17}_{-0.08}}~{\rm MeV}\\
\Gamma\left(\Xi_b^{\prime}({3\over2}^-)\to \Xi_b^{\prime}\rho\to\Xi_b^{\prime}\pi\pi\right)=5.6\times10^{-4}~{\rm MeV}
\end{array}$&$27.3~{^{+28.5}_{-14.2}}$&$\Xi_b(6227)^-$
\\ \cline{3-4} \cline{6-8}
&&$\Xi^\prime_b({5\over2}^-)$&$6.24\pm0.14$&&
$\begin{array}{c}
\Gamma\left(\Xi_b^{\prime}({5\over2}^-)\to \Xi_b\pi\right)=8.1~{^{+11.2}_{-~5.7}}~{\rm MeV}\\
\Gamma\left(\Xi_b^{\prime}({5\over2}^-)\to \Lambda_b K\right)=3.4~{^{+5.1}_{-2.2}}~{\rm MeV}\\
\Gamma\left(\Xi_b^{\prime}({5\over2}^-)\to \Xi_b^{\prime}\pi\right)=0.17~{^{+0.24}_{-0.11}}~{\rm MeV}\\
\Gamma\left(\Xi_b^{\prime}({5\over2}^-)\to \Xi_b^{*}\pi\right)=0.58~{^{+0.80}_{-0.38}}~{\rm MeV}\\
\Gamma\left(\Xi_b^{\prime}({5\over2}^-)\to\Xi_b^{*}\rho\to\Xi_b^{*}\pi\pi\right)=6.4\times10^{-5}~{\rm MeV}
\end{array}$
&$12.3~{^{+12.3}_{-~6.1}}$&--
\\ \hline \hline
\multirow{6}{*}{$\Omega_b$}&\multirow{2}{*}{$[\mathbf{6}_F,1,0,\rho]$}&$\Omega_b({1\over2}^-)$&$6.32 \pm 0.11$&\multirow{2}{*}{$2 \pm 1$}&--&$\sim~0$&\multirow{2}{*}{$\Omega_b(6316)^-$}
\\ \cline{3-4} \cline{6-7}
&&$\Omega_b({3\over2}^-)$&$6.32 \pm 0.11$&&--&$\sim~0$
\\ \cline{2-8}
&\multirow{2}{*}{$[\mathbf{6}_F,1,1,\lambda]$}&$\Omega_b({1\over2}^-)$&$6.34 \pm 0.10$&\multirow{2}{*}{$6 \pm 2$}&--&$\sim~0$&$\Omega_b(6330)^-$
\\ \cline{3-4} \cline{6-8}
&&$\Omega_b({3\over2}^-)$&$6.34 \pm 0.09$&&--&$\sim~0$&$\Omega_b(6340)^-$
\\ \cline{2-8}
&\multirow{2}{*}{$[\mathbf{6}_F,2,1,\lambda]$}&$\Omega_b({3\over2}^-)$&$6.35\pm0.13$&\multirow{2}{*}{$10\pm4$}&
$\begin{array}{c}
\Gamma\left(\Omega_b({3\over2}^-)\to \Xi_b K\right)=4.6~{^{+3.3}_{-1.9}}~{\rm MeV}
\end{array}$&$4.6~{^{+3.3}_{-1.9}}$&$\Omega_b(6350)^-$
\\ \cline{3-4} \cline{6-8}
&&$\Omega_b({5\over2}^-)$&$6.36\pm0.12$&&$\begin{array}{c}
\Gamma\left(\Omega_b({5\over2}^-)\to \Xi_b K\right)=2.5~{^{+3.5}_{-1.6}}~{\rm MeV}
\end{array}$ &$2.5~{^{+3.5}_{-1.6}}$&--
\\ \hline\hline
\end{tabular}
\label{tab:result}
\end{center}
\end{table*}

%
\section*{Acknowledgments}

We thank Er-Liang Cui and Qiang Mao for useful discussions.
This project is supported by
the National Natural Science Foundation of China under Grant No.~11722540
and
the Fundamental Research Funds for the Central Universities.

\appendix

\section{Sum rule equations}
\label{sec:othersumrule}

In this appendix we give several examples of sum rule equations, which are used to study $D$-wave decays of $P$-wave bottom baryons into ground-state bottom baryons and pseudoscalar mesons.

\begin{widetext}
The sum rule equation for the $\Xi_b^{\prime-}[{1\over2}^-]$ belonging to $[\mathbf{6}_F, 1 , 0, \rho]$ is
\begin{eqnarray}
&& G_{\Xi_b^{\prime-}[{1\over2}^-] \rightarrow \Xi_b^{*0}\pi^-} (\omega, \omega^\prime)
=  g_{\Xi_b^{\prime-}[{1\over2}^-] \rightarrow \Xi_b^{*0}\pi^-}\times{ f_{\Xi_b^-[{1\over2}^-]} f_{\Xi_b^{*0}} \over (\bar \Lambda_{\Xi_b^-[{1\over2}^-]} - \omega^\prime) (\bar \Lambda_{\Xi_b^{*0}} - \omega)}
\\ \nonumber &=& \int_0^\infty dt \int_0^1 du e^{i(1-u)\omega^\prime t} e^{iu\omega t}\times 4\times \Big(\frac{f_\pi m_s^2 m_\pi^2 u}{96(m_u+m_d)\pi^2}\phi_{3;\pi}^\sigma(u)+\frac{f_\pi m_s u}{128 \pi^2}\phi_{4;\pi}(u)
\\ \nonumber &&+\frac{f_\pi u}{24}\langle\bar s s\rangle\phi_{2;\pi}(u)+\frac{f_\pi m_s u}{8\pi^2 t^2}\phi_{2;\pi}(u)+\frac{f_\pi m_\pi^2 u}{24(m_u+m_d)\pi^2 t^2}\phi_{3;\pi}^\sigma(u)+\frac{f_\pi m_s m_\pi^2 u t^2}{576(m_u+m_d)}\langle\bar s s\rangle\phi_{3;\pi}^\sigma(u)
\\ \nonumber &&+\frac{f_\pi u t^2}{384}\langle \bar s s\rangle\phi_{4;\pi}(u)+\frac{f_\pi u t^2}{384}\langle g_s \bar s \sigma G s\rangle\phi_{2;\pi}(u)+\frac{f_\pi u t^4}{6144}\langle g_s \bar s \sigma G s\rangle\phi_{4;\pi}(u)\Big)
\\ \nonumber &-&\int_0^\infty dt \int_0^1 du \int \mathcal{D}\underline{\alpha}e^{i\omega^\prime t(\alpha_2+u\alpha_3)}e^{i\omega t(1-\alpha_2-u\alpha_3)}\times{1\over2}\times\Big(\frac{i f_{3\pi} u^2 \alpha_3}{4\pi^2 t v\cdot q}\Phi_{3;\pi}(\underline{\alpha})+\frac{i f_{3\pi} u \alpha_2 v \cdot q}{4\pi^2 t}\Phi_{3;\pi}(\underline{\alpha})
\\ \nonumber &&+\frac{i f_{3\pi}u \alpha_3 v \cdot q}{4\pi^2 t}\Phi_{3;\pi}(\underline{\alpha})-\frac{i f_{3\pi} u v \cdot q}{4\pi^2 t}\Phi_{3;\pi}(\underline{\alpha})+\frac{i f_{3\pi}\alpha_2 v \cdot q}{4\pi^2 t}\Phi_{3;\pi}(\underline{\alpha})-\frac{i f_{3\pi} v \cdot q}{4\pi^2 t}\Phi_{3;\pi}(\underline{\alpha})
\\ \nonumber &&-\frac{f_{3\pi} u}{4\pi^2 t^2}\Phi_{3;\pi}(\underline{\alpha})\Big) \, .
\end{eqnarray}

The sum rule equation for the $\Omega_b^-[{1\over2}^-]$ belonging to $[\mathbf{6}_F, 0, 1, \lambda]$ is
\begin{eqnarray}
G_{\Omega_b^-[{1\over2}^-] \rightarrow \Xi_b^{*0}K^-} (\omega, \omega^\prime)
=  g_{\Omega_b^-[{1\over2}^-] \rightarrow \Xi_b^{*0}K^-}\times{ f_{\Omega_b^-[{1\over2}^-]} f_{\Xi_b^{*0}} \over (\bar \Lambda_{\Omega_b^-[{1\over2}^-]} - \omega^\prime) (\bar \Lambda_{\Xi_b^{*0}} - \omega)}
= 0 \, .
\end{eqnarray}

The sum rule equation for the $\Sigma_b^-[{3\over2}^-]$ belonging to $[\mathbf{6}_F, 1, 1, \lambda]$ is
\begin{eqnarray}
&& G_{\Sigma_b^-[{3\over2}^-] \rightarrow \Sigma_b^0 \pi^-} (\omega, \omega^\prime)
=  g_{\Sigma_b^-[{3\over2}^-] \rightarrow \Sigma_b^0 \pi^-}\times{ f_{\Sigma_b^-[{3\over2}^-]} f_{\Sigma_b^0} \over (\bar \Lambda_{\Sigma_b^-[{3\over2}^-]} - \omega^\prime) (\bar \Lambda_{\Sigma_b^0} - \omega)}
\\ \nonumber &=& \int_0^\infty dt \int_0^1 du e^{i(1-u)\omega^\prime t}e^{iu\omega t}\times8\times \Big (\frac{i f_\pi u}{4\pi^2 t^3}\phi_{2;\pi}(u)+\frac{i f_\pi u}{64 \pi^2 t}\phi_{4;\pi}(u)
\\ \nonumber &&-\frac{i f_\pi m_\pi^2 u t}{144 (m_u+m_d)}\langle \bar q q\rangle\phi_{3;\pi}^\sigma(u)-\frac{i f_\pi m_\pi^2 u t^3}{2304(m_u+m_d)}\langle g_s \bar q \sigma G q\rangle\phi_{3;\pi}^\sigma(u)\Big)
\\ \nonumber &-&\int_0^\infty dt \int_0^1 du \int\mathcal{D}\underline{\alpha}e^{i\omega^\prime t(\alpha_2+u \alpha_3)}e^{i\omega t(1-\alpha_2-u \alpha_3)}\times\Big(\frac{i f_\pi u^2 \alpha_3}{8\pi^2 t}\Phi_{4;\pi}(\underline{\alpha})+\frac{i f_\pi u \alpha_2}{8\pi^2 t}\Phi_{4;\pi}(\underline{\alpha})
\\ \nonumber &&+\frac{i f_\pi u \alpha_3}{16\pi^2 t}\Phi_{4;\pi}(\underline{\alpha})+\frac{i f_\pi u\alpha_3}{16\pi^2 t}\widetilde \Phi_{4;\pi}(\underline{\alpha})-\frac{i f_\pi u}{8\pi t}\Phi_{4;\pi}(\underline{\alpha})+\frac{i f_\pi\alpha_2}{16\pi^2 t}\Phi_{4;\pi}(\underline{\alpha})+\frac{i f_\pi \alpha_2}{16 \pi^2 t}\widetilde\Phi_{4;\pi}(\underline{\alpha})
\\ \nonumber &&-\frac{i f_\pi}{16 \pi^2 t}\Phi_{4;\pi}(\underline{\alpha})-\frac{i f_\pi}{16 \pi^2 t}\widetilde\Phi_{4;\pi}(\underline{\alpha})+\frac{f_\pi u}{8\pi^2 t^2 v \cdot q}\Psi_{4;\pi}(\underline{\alpha})+\frac{3 f_\pi u}{8\pi^2 t^2 v \cdot q}\widetilde\Psi_{4;\pi}(\underline{\alpha})-\frac{f_\pi}{8\pi^2 t^2 v \cdot q}\Phi_{4;\pi}(\underline{\alpha})
\\ \nonumber &&-\frac{3 f_\pi}{8\pi^2 t^2 v \cdot q}\widetilde\Phi_{4;\pi}(\underline{\alpha})+\frac{f_\pi}{8\pi^2 t^2 v \cdot q}\Psi_{4;\pi}(\underline{\alpha})-\frac{f_\pi}{8\pi^2 t^2 v \cdot q}\widetilde\Psi_{4;\pi}(\underline{\alpha})\Big )\, .
\end{eqnarray}

The sum rule equation for the $\Sigma_b^-[{3\over2}^-]$ belonging to $[\mathbf{6}_F, 2, 1, \lambda]$ is
\begin{eqnarray}
&& G_{\Sigma_b^-[{3\over2}^-] \rightarrow \Lambda_b^0\pi^-} (\omega, \omega^\prime)
=  g_{\Sigma_b^-[{3\over2}^-] \rightarrow \Lambda_b^0\pi^-}\times{ f_{\Sigma_b^-[{3\over2}^-]} f_{\Lambda_b^0} \over (\bar \Lambda_{\Sigma_b^-[{3\over2}^-]} - \omega^\prime) (\bar \Lambda_{\Lambda_b^0} - \omega)}
\\ \nonumber &=& \int_0^\infty dt \int_0^1 du e^{i (1-u) \omega^\prime t} e^{i u \omega t} \times 8 \times \Big (\frac{f_\pi m_\pi^2 u}{12(m_u+m_d)\pi^2 t^2}\phi_{3;\pi}^\sigma(u)+\frac{f_\pi u t^2}{12}\langle\bar q q\rangle\phi_{2;\pi}(u)
\\ \nonumber &&+\frac{f_\pi u t^2}{192}\langle \bar q q\rangle\phi_{4;\pi}(u)+\frac{f_\pi u}{192}\langle g_s \bar q \sigma G q\rangle\phi_{2;\pi}(u)+\frac{f_\pi u t^4}{3072}\langle g_s \bar q \sigma G q\rangle\phi_{4;\pi}(u)\Big )
\\ \nonumber &-&\int_0^\infty dt \int_0^1 du\int\mathcal{D}\underline{\alpha}e^{i\omega^\prime t(\alpha_2+u\alpha_3)}e^{i\omega t(1-\alpha_2-u\alpha_3)}\times\Big(\frac{f_3\pi u}{2\pi^2 t^2}\Phi_{3;\pi}(\underline{\alpha})-\frac{f_{3\pi}}{2\pi^2 t^2}\Phi_{3;\pi}(\underline{\alpha})
\\ \nonumber &&+\frac{i f_{3\pi} u^2 \alpha_3 v \cdot q}{2\pi^2 t}\Phi_{3;\pi}(\underline{\alpha})+\frac{i f_{3\pi} u \alpha_2 v \cdot q}{2\pi^2 t}\Phi_{3;\pi}(\underline{\alpha})-\frac{i f_{3\pi} u v \cdot q}{2\pi^2 t}\Phi_{3;\pi}(\underline{\alpha})\Big) \, .
\end{eqnarray}

\end{widetext}

%

%

\end{document}